\documentclass[%
 reprint,
superscriptaddress,
 amsmath,amssymb,
 aps,
prl,
]{revtex4-2}

\usepackage{graphicx}% Include figure files
\usepackage{dcolumn}% Align table columns on decimal point
\usepackage{bm}% bold math
\usepackage{hyperref}% add hypertext capabilities
\usepackage{orcidlink}
\usepackage{xcolor}
\begin{document}

\title{Nuclear matter radii from molecular rotations using ultra-high-resolution {spectroscopy}}
% \title{Nuclear matter radii from molecular rotations using next-generation spectrometers}

\author{
    Michail Athanasakis-Kaklamanakis
    % \orcidlink{0000-0003-0336-5980}
}
\altaffiliation[Present address: ]{Centre for Cold Matter, Imperial College London, SW7 2AZ London, United Kingdom}
 \email{m.athkak@cern.ch}
\affiliation{Experimental Physics Department, CERN, CH-1211 Geneva 23, Switzerland}
\affiliation{KU Leuven, Instituut voor Kern- en Stralingsfysica, B-3001 Leuven, Belgium}

\author{Gerda Neyens}%\orcidlink{0000-0001-8613-1455}}
 \email{gerda.neyens@kuleuven.be}
\affiliation{KU Leuven, Instituut voor Kern- en Stralingsfysica, B-3001 Leuven, Belgium}

\date{\today}% It is always \today, today,

\begin{abstract}
The rotational constant parametrizes the relative spacing between a molecule's rotational energy levels. It depends on the molecule's classical moments of inertia, which, in all studies, are expressed by treating the constituent nuclei as point masses separated by the bond length. We point out that treating the finite nuclear size leads to a correction to the rotational constant at the Hz level, which is resolvable with recently developed ultra-high-resolution molecular spectrometers. Nuclear-model-independent measurements of nuclear matter radii can thus be {envisioned in the future using} such apparatuses, advancing beyond the existing hadronic scattering experiments and further developing the intersection of nuclear and molecular physics. {In the present time, it appears that the computational ability of \textit{ab initio} quantum chemistry might be the limiting factor to the technical readiness of the approach. To test the premises of the proposed method, we call for benchmark experiments using HD$^+$ that are feasible with state-of-the-art experiment and theory.}
\end{abstract}

\maketitle

\section{Introduction} \label{sec:intro}
Molecules are composed of two or more atoms joined by chemical bonds that couple the motion of the constituent nuclei. Even diatomic molecules, which have the simplest geometry, possess vibrational and rotational degrees of freedom. As a result, their spectra exhibit a vibrational structure for each electronic level, and a rotational structure within each vibrational sub-level of each electronic level.

The rotational spectra of diatomic molecules have been extensively studied for more than a century~\cite{Herzberg1945}, and it is well-understood that the spacing between successive lines in a rotational spectrum is intimately linked to the moments of inertia of the molecule. Invariably, the point-nucleus approximation is used to express the molecular moments of inertia and thus to extract the rotational constants~\cite{Herzberg1945,PaulingWilsonBook,BrownCarringtonBook,AtkinsPhysicalChemistry}. This approximation is certainly justified due to the large difference in scale between nuclear radii ($\sim$10$^{-15}$~m) and the internuclear distance in a typical diatomic molecule ($\sim$10$^{-10}$~m). Corrections due to treating the finite nuclear size can thus be expected to be extremely small, and so far considered negligible.

In the last few years, however, advances in microwave and infrared spectrometers with an unprecedented resolution~\cite{Satterthwaite2022} and absolute accuracy~\cite{Aiello2022} have been constructed, being capable of determining transition energies and thus rotational constants with sub-Hz precision and kHz accuracy. {Moreover, trap-based laser spectroscopy of light molecular ions has also demonstrated the control of statistical and systematic errors at the sub-kHz level~\cite{Alighanbari2020,Patra2020}.} Therefore, it is necessary to revisit the point-nucleus approximation in the molecular moment of inertia and consider the correction due to the full treatment of the finite nuclear size.

In light of these recent technical developments, here we point out that the emergent technology in molecular spectroscopy is sensitive to the matter radii of the constituent nuclei. We firstly provide the expression for the rotational constant in diatomic molecules where the constituent nuclei are treated as spheres of finite size. Given the resulting expression, we discuss how measurements of nuclear matter radii can be performed via ultra-high-resolution rotational spectroscopy of diatomic molecules with state-of-the-art and next-generation {spectroscopic techniques}. Compared to the {existing and established experimental methods} for measurements of {the nuclear matter radius}, the proposed approach offers 
{a nuclear-model-independent determination of nuclear matter radii, with potentially higher precision and significantly smaller isotope-dependent systematic uncertainties.}
% a high-precision and less costly pathway with significantly smaller, isotope-independent, and controlled systematic uncertainties.

\section{Molecular rotations}
\begin{figure}
    \centering
\includegraphics[width=0.45\textwidth]{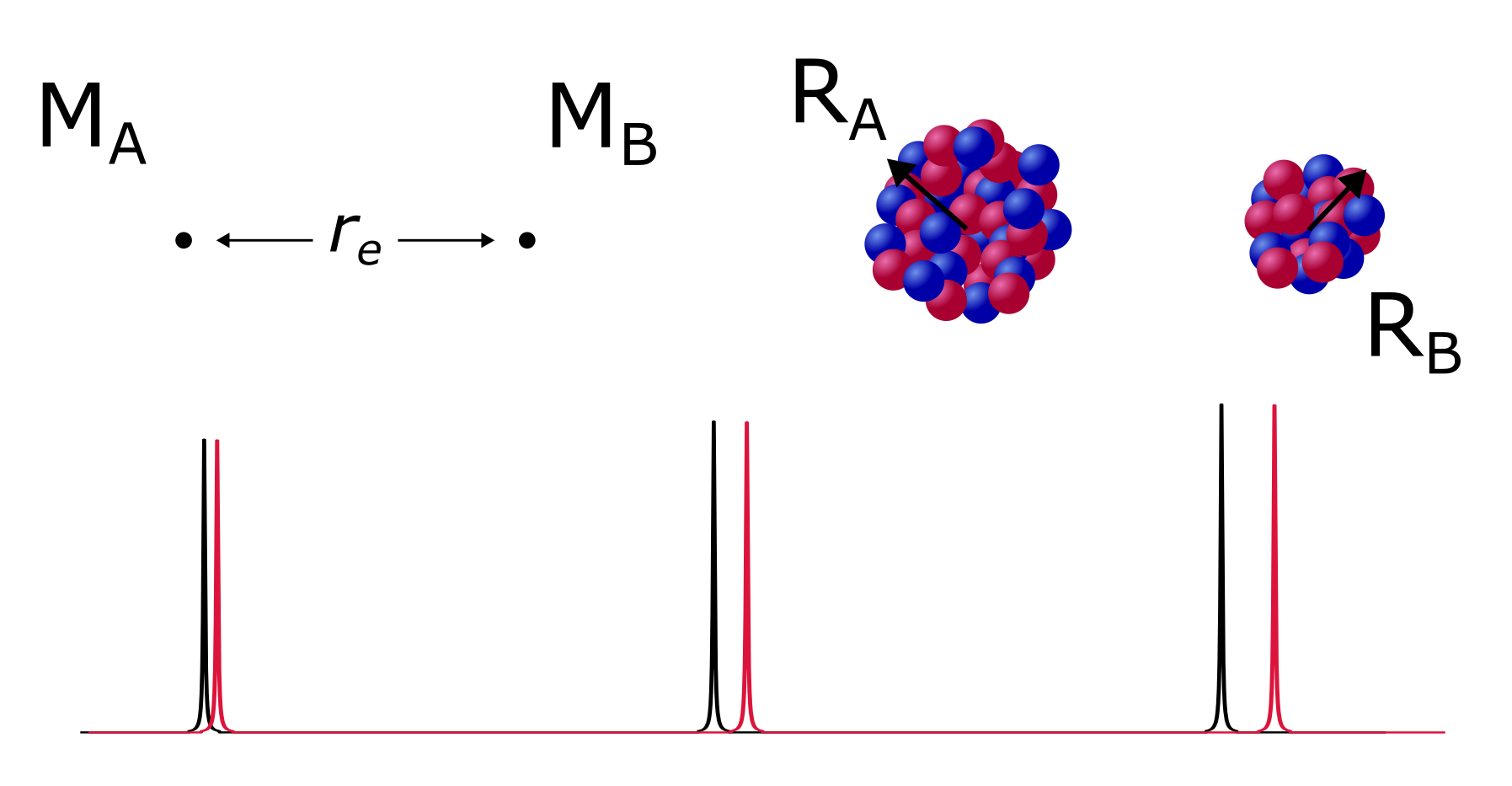}
    \caption{Top: Conceptual comparison between a point-nucleus (left) and a finite-nucleus (right) representation of a diatomic molecule. Bottom: Schematic effect of a change in the rotational constant on the relative spacing of three successive rotational lines.}
    \label{fig:main}
\end{figure}

Depending on the size of the diatomic molecule, pure rotational spectra, which are composed of transitions between rotational levels within the same electronic-vibrational state, are observed between the microwave and millimeter-wave regions of the electromagnetic spectrum. The rotational lines of a diatomic molecule, characterized by the rotational quantum number $J$, follow a characteristic, diverging pattern at the first order, which is given by the expression~\cite{Herzberg1945,PaulingWilsonBook,BrownCarringtonBook}:

\begin{equation} \label{eq:rotational_energy}
    E(J) = B_v J(J+1)
\end{equation}
where $B_v$ is called the rotational constant and it generally shows a dependence on the vibrational quantum number $v$. A schematic example of the effect of changes in the rotational constant on the spacing of rotational lines is shown in Fig.~\ref{fig:main}. The values of the rotational constant for different values of $v$ are related to an \textit{equilibrium} rotational constant $B_e$ through the expression~\cite{PaulingWilsonBook}:

\begin{equation} \label{eq:Bv_to_Be}
    B_v = B_e - \alpha_e \left( v + \frac{1}{2} \right)
\end{equation}
where $\alpha_e$ is a constant that depends on the parameters of the Morse potential in the vibrational term of the molecular Hamiltonian~\cite{PaulingWilsonBook}, and $B_e$ has the expression:

\begin{equation} \label{eq:B_e}
    B_e = \frac{h}{8\pi^2 c I_e}
\end{equation}

In Eq.~\ref{eq:B_e}, $I_e$ is the equilibrium moment of inertia. That is, the moment of inertia of the molecule assuming that the constituent nuclei are placed at a fixed distance from each other that is equal to the equilibrium bond length $r_e$.

In the point-nucleus approximation, the two nuclear masses $M_A$ and $M_B$ have an infinitely small radius and are separated by a distance $r_e$. Therefore, the point-nucleus moment of inertia is expressed as $I_e^p = \mu r_e^2$ where $\mu$ is the reduced molecular mass $\mu = \frac{M_A M_B}{M_A + M_B}$ and thus

\begin{equation} \label{eq:B_pointnucleus}
    B_e^p = \frac{h}{8\pi^2 c} \frac{1}{\mu r_e^2}
\end{equation}
is the rotational constant for the point-nucleus case.

To arrive at the molecular moment of inertia when treating the nuclei as spheres of radius $R_A$ and $R_B$, respectively, the parallel axis theorem can be used. Consider nucleus $A$ as placed at a distance $d_A$ from the center of mass of the molecule and nucleus $B$ as placed at a distance $d_B$. Due to the symmetry of a diatomic molecule, the molecular center of mass will lie along the internuclear axis, and by definition of the bond length, $r_e = d_A + d_B$.

The moment of inertia of a solid sphere about any axis going through its center is:
\begin{equation} \label{eq:MoI_sphere}
    I_{\rm{sphere}}=\frac{2}{5}M R^2
\end{equation}
where $M$ is the sphere's mass and $R$ its radius. From the parallel axis theorem, the moment of inertia of a solid sphere about an axis at a distance $d$ from its center is:
\begin{equation} \label{eq:parallel_axis}
    I = I_{\rm{sphere}} + M d^2
\end{equation}

For two nuclei rotating about the same axis in the same direction, the total moment of inertia of the molecule is equal to the sum of the moments of inertia of the nuclei. Therefore, using Eqs.~\ref{eq:MoI_sphere} and \ref{eq:parallel_axis}, the moment of inertia of the diatomic molecule is:
\begin{equation*}
    I_e^s = \frac{2}{5}M_A R_A^2 + M_A d_A^2 + \frac{2}{5}M_B R_B^2 + M_B d_B^2
\end{equation*}
\begin{equation} \label{eq:I_mol_1}
    \Leftrightarrow I_e^s = \frac{2}{5} \left( M_A R_A^2 + M_B R_B^2 \right) + M_A d_A^2 + M_B d_B^2
\end{equation}

The last two terms are equivalent to the point-nucleus moment of inertia $I_e^p$. This can be seen by considering the molecule's center of mass as defining the zero coordinate of the internuclear axis. The locations of nuclei $A$ and $B$ are thus:
\begin{equation}
    d_A = -\frac{M_B r_e}{M_A + M_B}
\end{equation}

\begin{equation}
    d_B = \frac{M_A r_e}{M_A + M_B}
\end{equation}

Therefore:
\begin{multline*}
    M_A d_A^2 + M_B d_B^2 = \frac{M_A M_B^2 r_e^2}{(M_A + M_B)^2} + \frac{M_B M_A^2 r_e^2}{(M_A + M_B)^2} 
    \\ 
    = \frac{M_A M_B M_B + M_B M_A M_A }{(M_A + M_B)^2} r_e^2 
    \\
    = \frac{M_A M_B (M_A + M_B)}{(M_A + M_B)^2} r_e^2 
    = \frac{M_A M_B}{M_A + M_B} r_e^2
    \end{multline*}

And thus:
\begin{equation}
   M_A d_A^2 + M_B d_B^2 = \mu r_e^2
\end{equation}

Eq.~\ref{eq:I_mol_1} is therefore equal to:
\begin{equation} \label{eq:I_mol_2}
    I_e^s = \frac{2}{5} \left( M_A R_A^2 + M_B R_B^2 \right) + I_e^p
\end{equation}

At the point-nucleus limit, where $R_A=R_B=0$, $I_e^s = I_e^p$, as expected. The equilibrium rotational constant for a diatomic molecule can thus be expressed as:
\begin{equation} \label{eq:B_e_final}
    B_e = \frac{h}{8\pi^2 c} \frac{1}{ \frac{2}{5} \left( M_A R_A^2 + M_B R_B^2 \right) + \mu r_e^2}
\end{equation}

\section{Nuclear matter radii}
\begin{figure}
    \centering
    \includegraphics[width=0.5\textwidth]{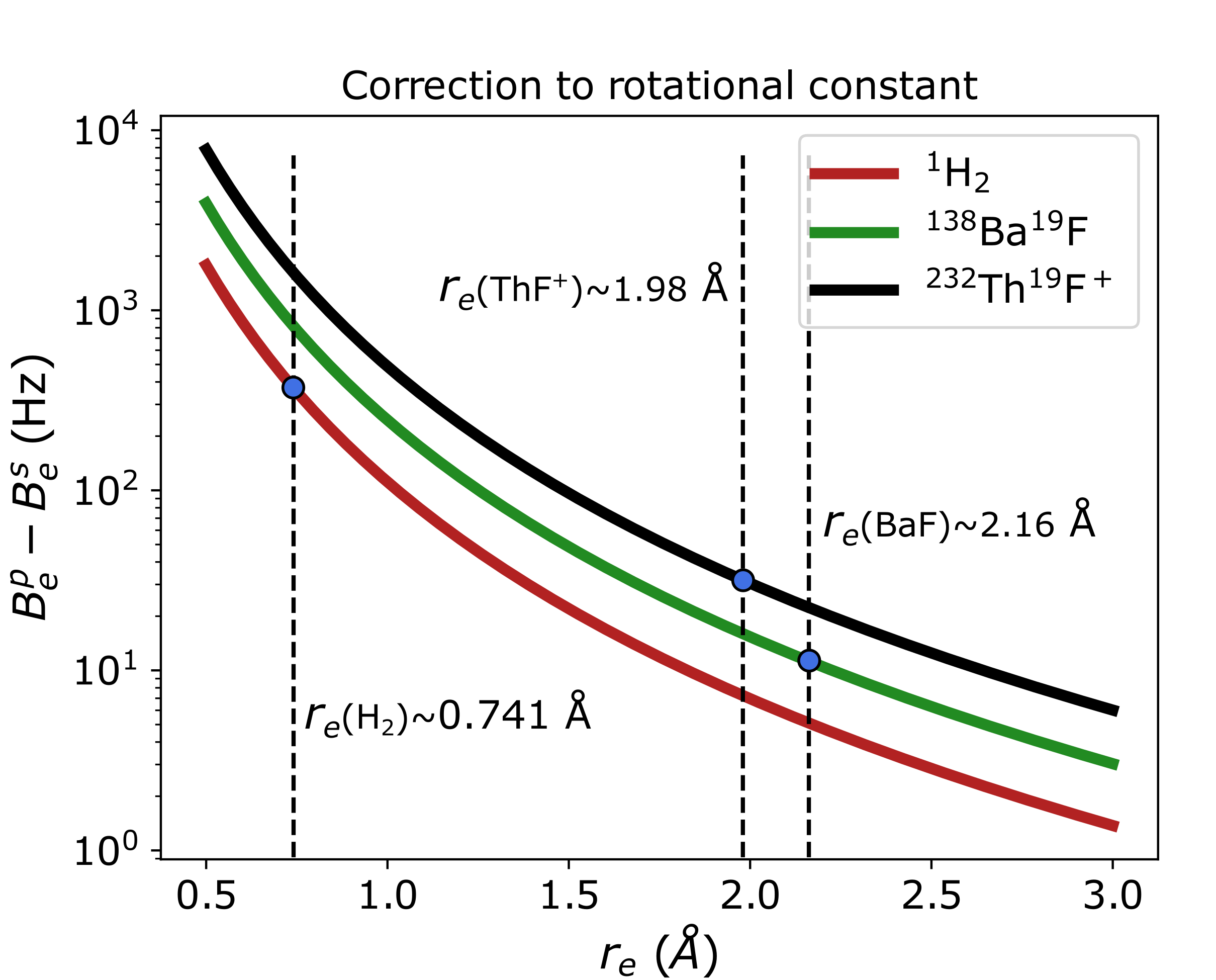}
    \caption{Calculation of correction (in Hz) to the rotational constant due to treating the finite size of the nuclei as opposed to the point-nucleus approximation. The nuclear matter radii used to calculate the corrections were approximated as equal to the charge radii and taken from Ref.~\cite{Angeli2013}. Nuclear masses were taken from the 2020 Atomic Mass Evaluation~\cite{AME2020} and equilibrium bond lengths were taken from the NIST Chemistry WebBook.}
    \label{fig:correction}
\end{figure}

As per Eq.~\ref{eq:I_mol_2}, the moment of inertia of a diatomic molecule when treating the finite nuclear size has a correction that depends on the constituent nuclear masses and radii. Fig.~\ref{fig:correction} shows how the correction scales as a function of the equilibrium bond length $r_e$ for three diatomic molecules of different mass. Evidently, the correction to the rotational constant, quantified as the difference between the point-nucleus and the finite-nucleus rotational constants $B_e^p$ and $B_e^s$, is in the order of a few tens of Hz for both the lightest possible molecule (H$_2$) and the actinide molecule ThF$^+$.

With sub-kHz resolution, the state of the art in {molecular} spectroscopy is already capable of determining the rotational constants of diatomic molecules with a Hz- or sub-Hz-level uncertainty, and further improvements can be expected in the future. Therefore, it is worthwhile considering how such a precision can be utilized to measure nuclear matter radii.

The size of a nucleus is most often discussed in terms of two observables; the mean-squared nuclear charge radius $\langle r^2 \rangle_{\rm{ch}}$, which is determined solely by the proton distribution, and the mean-squared nuclear matter radius $\langle r^2 \rangle_{\rm{m}}$, which is determined by both the proton and the neutron distributions.

The charge radius can be measured through a variety of complementary experimental techniques. Muonic-atom X-ray spectroscopy~\cite{Engfer1974} and electron scattering experiments~\cite{Hofstadter1956,deVries1987} have provided measurements of the absolute nuclear charge radii of stable isotopes for decades~\cite{Angeli2013}. Laser spectroscopy, on the other hand, provides high-precision measurements of the relative change in root-mean-squared (rms) nuclear charge radius between isotope pairs in a nuclear-model-independent manner, through the isotope shift~\cite{King2013}. By combining absolute measurements in stable isotopes and relative measurements in radioactive isotopes, the charge radii of a great amount of nuclei far from nuclear stability have been measured~\cite{Yang2023}. The study of nuclear charge radii has thus become a cornerstone of modern nuclear structure research, being pursued to investigate the emergence of new magic numbers~\cite{GarciaRuiz2016,Koszorus2021}, the disappearance and inversion of odd-even staggering~\cite{deGroote2020b,Verstraelen2019a}, signatures of rare nuclear effects~\cite{Miller2019}, and more. Recently, muonic-atom and electron-scattering experiments on radioactive isotopes were demonstrated for the first time~\cite{Adamczak2018,SCRIT2023}, opening a new pathway towards absolute charge radius determination in unstable nuclei.

Measurements of the nuclear matter radius~\cite{Brissaud1972,Alkhazov1978}, on the other hand, are significantly more challenging. Since the pioneering experiments by Rutherford and his group~\cite{Rutherford1911}, charged-particle scattering experiments have undergone continuous development and to this day are the most widespread approach to measuring the distribution of nuclear matter~\cite{Batty1989}, together with matter-antimatter annihilation approaches~\cite{Aumann2022}. Extracting the nuclear matter radius from hadronic scattering requires modeling of the underlying hadron dynamics, knowledge of which remains incomplete. Therefore, nuclear matter radii from hadronic scattering have large systematic uncertainties that are isotope-dependent and mostly uncontrolled~\cite{Thiel2019}. So far, only the PREX-2~\cite{PREX2021} experiment has come close to extracting the nuclear matter radius at the 1$\%$ level in statistical and systematic uncertainties, based on the determination of the neutral weak form factor from elastic electron scattering, thus avoiding the uncontrolled systematics of hadronic models.

As a result, while knowledge of the nuclear matter radius and the neutron distribution is highly important in nuclear physics, for instance to explore the formation of neutron skins~\cite{PREX2021}, the equation of state in neutron stars~\cite{Lattimer2000,Reed2021}, the structure of nuclear halos~\cite{Egelhof2002,Dobrovolsky2006}, and the limits of the nuclear landscape~\cite{Erler2012,Tsunoda2020}, their experimental study requires large-scale and costly reaction spectrometers at specialized facilities. Accessing the nuclear matter radius via ultra-high-resolution rotational spectroscopy of diatomic molecules thus offers an alternative experimental tool of great value, with a reduced cost, potential for lower statistical uncertainties, and controlled, isotope-independent systematic uncertainties.

The moment of inertia in Eq.~\ref{eq:I_mol_2} defines an equilibrium rotational constant $B_e$ (Eq.~\ref{eq:B_e_final}) for a diatomic molecule comprised of realistic, finite-size nuclei. The equilibrium rotational constant, however, is not directly measurable in the laboratory. Instead, the vibrational-state-dependent constants $B_v$ are measured, which are related to $B_e$ via Eq.~\ref{eq:Bv_to_Be}.

The constant $\alpha_e$ in Eq.~\ref{eq:Bv_to_Be} is related to the parameters of the internuclear potential used in the molecular Hamiltonian. For the {most commonly used} Morse potential, $\alpha_e$ depends on the three Morse potential parameters $\alpha$, $D$, and $r_e$ as~\cite{PaulingWilsonBook}:
\begin{equation} \label{eq:alpha_e}
    \alpha_e = \frac{3 h^2 \omega_e}{16 \pi^2 \mu r_e ^2 D} \left( \frac{1}{\alpha r_e} - \frac{1}{\alpha^2 r_e^2} \right) 
\end{equation}
where $\omega_e = \frac{\alpha}{2 \pi c} \sqrt{\frac{2D}{\mu}}$ is the harmonic frequency of the electronic state.

Consider a diatomic molecule with atoms $A$ and $B$ for which the atomic masses of both nuclei are known with high precision and accuracy (a precision of $10^{-9}$ atomic mass units or better~\cite{AME2020}), and the nuclear matter radius for atom $B$ has been measured in a scattering experiment. Combining Eqs.~\ref{eq:Bv_to_Be}, \ref{eq:B_e_final}, and \ref{eq:alpha_e}, it is seen that for each value of $v$, an equation with $R_A$, $\alpha$, $D$, and $r_e$ as unknown is constructed.

Therefore, a value for $R_A$ can be extracted if $B_v$ is measured for {enough} different values of $v$. {For the Morse potential and alternative functions~\cite{Wang2012Vibr}, which are described by three parameters, $B_v$ would have to be measured for at least $v=0-3$}. With four values of $B_v$, a set of four equations and four unknowns is formed. Adding further measurements of $B_v$ for higher values of $v$ will help reduce the uncertainty in the extracted parameters by constraining the linear form of Eq.~\ref{eq:Bv_to_Be}. While spectroscopy of excited vibrational states is not typical for cryogenic buffer-gas-cooled spectrometers, vibrational relaxation through buffer-gas collisions is significantly more inefficient than rotational and translational cooling~\cite{Hutzler2011}, which makes such spectrometers compatible with the proposed approach. {For ion trap experiments, excited vibrational states can be studied via optical or infrared excitations of overtone vibrational transitions.

A significant aspect of the approach proposed here is its versatility, as practically no electronic-state requirements are imposed on the system or technique chosen for the measurement of the rotational constant. As a result, the measurements can be performed in either neutral or ionic systems, in the ground or metastable states.}

\section{Higher-order corrections}
{Beyond the correction due to the finite nuclear size considered here, other corrections to the molecular rotational constant are to be expected at the level of precision that is necessary to extract the nuclear matter radius, such as quantum electrodynamics (QED) corrections. Hyperfine effects and the influence of the nuclear charge radius on the bond length may also need to considered at this level of precision.

The presently described method proposes extracting the nuclear matter radius by determining $B_e$ with very high precision, and then using the relation between the moment of inertia and $B_e$ to extract the matter radius. The extraction of $B_e$ can be done by measuring $B_v$ for several vibrational states of the electronic ground state, and then performing a fit as per Eq.~\ref{eq:Bv_to_Be}. In a fit of $B_v$ versus $\left( v + \frac{1}{2} \right)$, the y-intercept gives $B_e$ and the slope is related to the parameters of the vibrational potential used in the molecular Hamiltonian. One of these parameters is the equilibrium bond length $r_e$, knowledge of which is also necessary to use Eq.~\ref{eq:B_e_final} and extract the nuclear matter radii. Depending on the vibrational potential used in the Hamiltonian, $r_e$ could be extracted from the fit's slope.

So far it has been considered that if $B_v$ is measured for a sufficient number of vibrational states with high precision, then $B_e$ and $r_e$ could be extracted and provide the nuclear matter radius of interest without the need to calculate any properties of the molecule with \textit{ab initio} quantum chemistry. However, at the level of precision discussed here, there are several points that need consideration.

Most importantly, it is not clear to what degree the relation between $B_e$ and $B_v$ in Eq.~\ref{eq:Bv_to_Be} is exact at this level -- in fact, it is certainly not fully linear. The description of this relation might depend to some extent on the potential energy function used, and it is known that the most commonly used Morse potential energy function for diatomic molecules has limited accuracy, leading to the development of more elaborate potential energy functions (see Refs.~\cite{DelSolMesa1998,Wang2012Vibr,AbuShady2023} and references therein). However, none of these potentials have been held to the test of ultra-high-precision spectroscopy so far.

Therefore, the validity of Eq.~\ref{eq:Bv_to_Be} at the precision required by this method remains to be confirmed. It is expected that additional terms will need to be considered in Eq.~\ref{eq:Bv_to_Be} at higher order of $\left( v + \frac{1}{2} \right)$. These corrections would make Eq.~\ref{eq:Bv_to_Be} non-linear, but would not affect the extraction of $B_e$ through the y-intercept of a fit between $B_v$ and $\left( v + \frac{1}{2} \right)$; the higher-order terms do not need to be understood at an analytical, first-principles level to extract $B_e$ from the y-intercept. Nevertheless, these terms will impact the ability to extract $r_e$, and theoretical and experimental investigations into the applicability of Eq.~\ref{eq:Bv_to_Be} at very high precision are still necessary.

Most likely, extracting the nuclear matter radius from $B_e$ will require at least partly input from \textit{ab initio} calculations; for instance, via a highly precise calculation of $r_e$. In this case, further corrections need to be considered, such as the influence of the nuclear charge radii on $r_e$, as well as radiative and relativistic corrections.}

{Especially for the correction due to the finite nuclear charge radius, there are indications that the correction can be at the kHz level or higher.} Within the Dunham parametrization of the rovibrational energy levels of diatomic molecules~\cite{Dunham1932}, the rotational constant $B$ is expressed as the $k,l=0,1$ Dunham parameter $Y_{01}$. While early studies considered that the Dunham parameters need only be scaled by the reduced molecular mass upon isotopic substitution of the constituent nuclei~\cite{Dunham1932,Herzberg1945}, it has long been noticed that this relation is incomplete and instead the nuclear mass and nuclear charge radius need to be taken into account in an explicit manner~\cite{Schlembach1982,Knecht2012}. Recently, we showed that the nuclear radius and mass corrections to the Dunham expansion can be viewed as the molecular equivalent of the more familiar atomic field and mass shifts and utilized directly in a King plot with atomic data~\cite{AthanasakisKaklamanakis2023kingplot}.

Despite the extensive evidence of the validity of the nuclear mass and field corrections to the Dunham parameters, however, they are not routinely considered in the expression of the molecular rotational constant in spectroscopic studies. Instead, their influence is typically "absorbed" into an effective equilibrium bond distance that is extracted from the measured spectra using the definition in Eq.~\ref{eq:B_pointnucleus}. With current and planned advances in the accuracy and precision of state-of-the-art spectroscopic equipment, these corrections need to be taken into consideration in a systematic and consistent manner. Especially for the accurate extraction of nuclear matter radii, as proposed here, an accurate value for the equilibrium bond length $r_e$ needs to be determined, and these corrections must therefore be fully treated. Further corrections beyond the nuclear mass and field shifts also need to be explored.

{To assess the magnitude of these corrections to $r_e$, \textit{ab initio} molecular theory has to be employed. However, the correction to the rotational constant due to the matter radius decreases for heavier molecules (Fig.~\ref{fig:correction}), and typically so does the computational precision of quantum chemistry as well. Meanwhile, higher-order corrections to the bond length might increase in significance for heavier molecules, such as the correction due to the nuclear charge radius and QED effects.

While QED corrections to electronic transition energies have been calculated for heavy molecules, even up to the theoretical superheavy molecule E(120)F~\cite{Skripnikov2021b}, the computational accuracy and precision remain limited considering the requirements of our proposed method. Progress in \textit{ab initio} relativistic calculations relevant to multi-electron systems is ongoing, 
% such as reaching an agreement better than 99.5$\%$ for the excitation energies of RaF even at high excitation energy~\cite{AthKak2023Excited}, 
but the computational precision and accuracy are still many orders of magnitude from the level required by the method presented here.

As a result, only light molecules appear accessible by the currently proposed method in the present time. This does not eliminate the near-term nuclear-physics impact of the proposed method, since the accurate and precise determination of the nuclear matter radius in light nuclei can be used as a benchmark of a large number of emerging \textit{ab initio} nuclear theory techniques~\cite{Navratil2016,Hergert2020}, whose applicability is often limited to light nuclei. This research potential also includes direct benchmarks of large-scale lattice quantum chromodynamics calculations using the structure of light nuclei~\cite{Beane2014NPLQCD,Chang2015NPLQCD,Parreno2021NPLQCD}.}

\section{Proposed cases}
\subsection{HD$^+$}
Extracting the nuclear matter radius via the method proposed here introduces a systematic uncertainty depending on the values used for the matter radii of the other nuclei in the molecule. As a result, diatomic molecules are highly preferable. Specifically, monohydrides are ideal, since the nuclear matter and charge radii of $^{1}$H are identical,
% . The value of the proton radius used in Eq.~\ref{eq:B_e_final} can be extracted with high precision and accuracy,
and progress is ongoing towards resolving the proton radius puzzle~\cite{Gao2022ProtonRadius}.

% Therefore, a natural first case to test the proposed method is molecular hydrogen, H$_2$, as the simplest consistency check could take the form of confirming that the two nuclear matter radii involved in Eq.~\ref{eq:B_e_final} are identical. 
{The hydrogen compounds remain the subject of several high-precision experimental and theoretical studies~\cite{Koelemeij2007,Bressel2012,Dickenson2013,Biesheuvel2016a,Biesheuvel2016b,Korobov2017,Aznabayev2019,Patra2020,Fink2020,Rau2020,Alighanbari2020,Germann2021,Kortunov2021,Alighanbari2023}. HD$^+$, in particular, possessing only one electron, is an ideal probe to test the quantum-mechanical description of a three-body system at the part-per-trillion level~\cite{Alighanbari2020,Patra2020,Germann2021}, without electron correlation effects. As a result, it also provides an excellent system to investigate the magnitude of the higher-order corrections required for the reliable extraction of nuclear matter radii.  As per Fig.~\ref{fig:correction}, the finite-radius correction in H$_2$ is expected to be in the order of 10$^2$~Hz, owing to the short bond length of the system, and so the nuclear matter radius should be accessible with the precision already demonstrated with optical spectroscopy.

% In terms of high-precision investigations of the potential energy function, 
Major breakthroughs have been achieved in the last decade on the vibrational-rotational structure of HD$^+$~\cite{Koelemeij2007,Dickenson2013,Patra2020,Germann2021,Kortunov2021,Alighanbari2023}, including high-precision investigations of the potential energy surface in the hydrogen compounds. The dissociation energy of molecular hydrogen was the first observable for which QED effects were explicitly calculated in a molecule~\cite{Komasa2011}. More recently, the theoretical value for the dissociation energy of H$_2$~\cite{Puchalski2019} was successfully benchmarked by a highly precise measurement that achieved an uncertainty of less than 5~kHz~\cite{Hoelsch2019}.

For several of the higher-order corrections mentioned in the previous section, results have already been reported for HD$^+$. In their recent work on the precision spectroscopy of HD$^+$~\cite{Alighanbari2020}, Alighanbari \textit{et al.} provided an uncertainty breakdown for the \textit{ab initio} calculation of a rotational transition frequency in HD$^+$ that included QED and finite-size corrections and reached a theoretical uncertainty of 18~Hz. These \textit{ab initio} calculations were successfully benchmarked for several hyperfine components of the rotational transition, confirming the validity of the presented theoretical results at the $\sim$10$^1$-Hz level. 

Importantly, the relativistic and finite-charge-radii corrections to the rotational transition frequency in HD$^+$ were calculated to be the most dominant, having a relative contribution of 10$^{-5}$ to the total transition frequency~\cite{Alighanbari2020}. For a larger system, containing a heavier nucleus with a larger charge and charge radius, both corrections would be expected to increase in absolute and relative magnitude even further, and thus their accurate calculation is critical.

Corrections of even higher order than those considered by Alighanbari \textit{et al.}~\cite{Alighanbari2020}, that is, going beyond three-loop radiative corrections, can be expected to contribute less than 100~Hz to the rotational transition frequency, while the finite-matter-radius correction to the rotational constant is estimated to be a few hundred Hz (see Fig.~\ref{fig:correction}). As the quantity of interest for our method is the rotational constant $B_e$, rather than individual rotational transition frequencies, the theoretical uncertainty of interest can be expected to be even better than the 18~Hz reported in Ref.~\cite{Alighanbari2020}. The reason is that corrections that shift individual rotational transitions the same way will be significantly suppressed when determining the rotational constant, as the latter is determined from the relative difference between rotational transitions (see for example Refs.~\cite{Dulick1998LiH,Markus2016OH+}, where order-of-magnitude higher precision is achieved for the rotational constant compared to the individual transition frequencies for light molecules).

Refs.~\cite{Alighanbari2020,Patra2020,Alighanbari2023} also demonstrate the control of systematic shifts in the spectroscopy of rotational and rovibrational lines to the sub-kHz level. These shifts are likely to also be suppressed when translated into a rotational-constant uncertainty, allowing the matter radius to be extracted without the further need for an order of magnitude improvement in the control of systematics. However, this has to be investigated directly. It is also important to note that systematic shifts were controlled to this level in ion trap experiments, and a similar level of control remains to be demonstrated in the more universally applicable cryogenic buffer-gas cell approaches of Refs.~\cite{Satterthwaite2022,Aiello2022}.

Overall, with statistical and systematic uncertainties already under control at the required level, both in experiment and theory, a campaign could be pursued with HD$^+$ to measure multiple overtone rovibrational transitions to different upper vibrational states, and determine $B_v$ for a range of $v$ (for instance, such measurements were performed in Ref.~\cite{McKellar1976} for neutral HD, but at a much lower precision than required for this method). These measurements can then be used to test the linearity of $B_v$ as a function of $\left( v + \frac{1}{2} \right)$ at the Hz level, and test the validity of Eq.~\ref{eq:Bv_to_Be}. As a next step, the consistency of Eq.~\ref{eq:B_e_final} can be tested with this molecule to extract the matter radius of the deuteron and compare it to the literature value from reaction experiments and nucleon-nucleon potential calculations~\cite{Wong1994DeuteronMatterRadius}.
}

\subsection{Calcium radii}
{Once the premises of the proposed method are tested with the spectroscopy of HD$^+$}, a measurement of the nuclear matter radii of $^{40,48}$Ca could also be pursued. The calcium isotopic chain ($Z=20$) is of central interest in nuclear structure research, as it has a magic proton number and two stable isotopes with a magic neutron number (at $N=20,28$), thus being doubly magic. The nuclear ground states of the doubly magic $^{40,48}$Ca are more energetically stable compared to nuclei in their vicinity, as seen through their binding energy, two-neutron separation energy, and transition energy and cross-section to their first excited states~\cite{CastenBook}. These observations have been acknowledged since the early development of the nuclear shell model and they form a staple of nuclear physics knowledge to this day.

Despite an excess of 8 neutrons, the nuclear charge radius of $^{48}$Ca is astonishingly similar to that of $^{40}$Ca~\cite{GarciaRuiz2016}. This is surprising and indicates either a deviation from the constancy of nucleon density within the nucleus as observed throughout the nuclear chart~\cite{CastenBook}, or much more likely the formation of a thick neutron skin in $^{48}$Ca.

To explore these conclusions, knowledge of the nuclear matter radii is necessary, which have been measured with proton elastic scattering~\cite{Igo1979,McCamis1986,Zenihiro2018} and collisions with a carbon target~\cite{Tanaka2000}. While different studies point to the formation of a considerably thicker neutron skin in $^{48}$Ca compared to $^{40}$Ca, there is noticeable statistical scatter between the different values reported for the rms nuclear matter radii of $^{40,48}$Ca that originates from the different nuclear-model assumptions employed in each analysis. On the other hand, the measured electric dipole polarizability of $^{48}$Ca implies a small neutron skin~\cite{Birkhan2017}. Lastly, Tanaka \textit{et al.}~\cite{Tanaka2000} reported an abrupt and prominent increase in the rms nuclear matter radius beyond $N=28$, which requires an independent validation.

{To extract the matter radius of calcium nuclei using the current method, calcium monohydride (CaH) would be the most reasonable choice due to the precise knowledge of the proton radius. CaH, which is the lightest possible calcium molecule, is significantly heavier than the hydrogen compounds discussed above. As a result, it cannot be reasonably expected that the theoretical precision achieved for HD$^+$ can also be achieved for CaH in the present time. We thus also call for the development of precision calculations of this and other light calcium molecules, to assess the readiness level of applying the currently proposed method to CaH and other calcium molecules for the extraction of nuclear matter radii.
}

\section{Conclusion}
By including the finite nuclear size in the expression of the equilibrium rotational constant, we point out that ultra-high-resolution molecular {spectroscopy techniques} are already sensitive to the influence of nuclear matter radii on the rotational structure of diatomic molecules. {With further improvements in calculations of the molecular bond length, ultra-high-resolution molecular spectroscopy can offer an alternative route to measuring nuclear matter radii, at a much reduced cost and with lower systematic uncertainties as compared to hadronic scattering experiments.
% With further improvements expected in the future, ultra-high-resolution molecular spectroscopy can be pursued for the measurement of nuclear matter radii with reduced cost and systematic uncertainties compared to hadronic scattering experiments.

The main limitation of the proposed method appears to be the precision and accuracy of \textit{ab initio} molecular calculations that might be required to determine the equilibrium bond length $r_e$, which is involved in the expression that relates the nuclear matter radii and the equilibrium rotational constant $B_e$. The necessary precision and accuracy of such calculations has been demonstrated in the case of HD$^+$, and achieving similar performance in calculations of heavier systems is necessary.

To test the validity of our proposal at the required level of precision, we propose that high-precision spectroscopy of HD$^+$ is pursued. Measurements and \textit{ab initio} calculations at the part-per-trillion level have already been achieved for this molecule, and existing experimental setups can be used to test the premises of our proposed method. We also propose the spectroscopy of CaH to determine the nuclear matter radii of $^{40,48}$Ca as a case of high nuclear physics importance~\cite{Tanaka2000,Birkhan2017}.
}

\section*{Acknowledgments}
\begin{acknowledgments}
{We would like to thank Alexander A. Breier (Technische Universit\"{a}t Berlin) and Shane G. Wilkins (Massachusetts Institute of Technology) for useful discussions. 

Financial support from FWO, as well as from the Excellence of Science (EOS) programme (No. 40007501) and the KU Leuven project C14/22/104, is acknowledged.}
\end{acknowledgments}

\bibliography{ref_mak}% Produces the bibliography via BibTeX.

\begin{thebibliography}{75}%
\makeatletter
\providecommand \@ifxundefined [1]{%
 \@ifx{#1\undefined}
}%
\providecommand \@ifnum [1]{%
 \ifnum #1\expandafter \@firstoftwo
 \else \expandafter \@secondoftwo
 \fi
}%
\providecommand \@ifx [1]{%
 \ifx #1\expandafter \@firstoftwo
 \else \expandafter \@secondoftwo
 \fi
}%
\providecommand \natexlab [1]{#1}%
\providecommand \enquote  [1]{``#1''}%
\providecommand \bibnamefont  [1]{#1}%
\providecommand \bibfnamefont [1]{#1}%
\providecommand \citenamefont [1]{#1}%
\providecommand \href@noop [0]{\@secondoftwo}%
\providecommand \href [0]{\begingroup \@sanitize@url \@href}%
\providecommand \@href[1]{\@@startlink{#1}\@@href}%
\providecommand \@@href[1]{\endgroup#1\@@endlink}%
\providecommand \@sanitize@url [0]{\catcode `\\12\catcode `\$12\catcode
  `\&12\catcode `\#12\catcode `\^12\catcode `\_12\catcode `\%12\relax}%
\providecommand \@@startlink[1]{}%
\providecommand \@@endlink[0]{}%
\providecommand \url  [0]{\begingroup\@sanitize@url \@url }%
\providecommand \@url [1]{\endgroup\@href {#1}{\urlprefix }}%
\providecommand \urlprefix  [0]{URL }%
\providecommand \Eprint [0]{\href }%
\providecommand \doibase [0]{https://doi.org/}%
\providecommand \selectlanguage [0]{\@gobble}%
\providecommand \bibinfo  [0]{\@secondoftwo}%
\providecommand \bibfield  [0]{\@secondoftwo}%
\providecommand \translation [1]{[#1]}%
\providecommand \BibitemOpen [0]{}%
\providecommand \bibitemStop [0]{}%
\providecommand \bibitemNoStop [0]{.\EOS\space}%
\providecommand \EOS [0]{\spacefactor3000\relax}%
\providecommand \BibitemShut  [1]{\csname bibitem#1\endcsname}%
\let\auto@bib@innerbib\@empty
%</preamble>
\bibitem [{\citenamefont {Herzberg}(1945)}]{Herzberg1945}%
  \BibitemOpen
  \bibfield  {author} {\bibinfo {author} {\bibfnamefont {G.}~\bibnamefont
  {Herzberg}},\ }\href@noop {} {\emph {\bibinfo {title} {{Molecular spectra and
  molecular structure}}}}\ (\bibinfo  {publisher} {D. van Nostrand},\ \bibinfo
  {year} {1945})\BibitemShut {NoStop}%
\bibitem [{\citenamefont {Pauling}\ and\ \citenamefont
  {Wilson}(2012)}]{PaulingWilsonBook}%
  \BibitemOpen
  \bibfield  {author} {\bibinfo {author} {\bibfnamefont {L.}~\bibnamefont
  {Pauling}}\ and\ \bibinfo {author} {\bibfnamefont {E.~B.}\ \bibnamefont
  {Wilson}},\ }\href@noop {} {\emph {\bibinfo {title} {{Introduction to quantum
  mechanics with applications to chemistry}}}}\ (\bibinfo  {publisher} {Courier
  Corporation},\ \bibinfo {year} {2012})\BibitemShut {NoStop}%
\bibitem [{\citenamefont {Brown}\ and\ \citenamefont
  {Carrington}(2003)}]{BrownCarringtonBook}%
  \BibitemOpen
  \bibfield  {author} {\bibinfo {author} {\bibfnamefont {J.~M.}\ \bibnamefont
  {Brown}}\ and\ \bibinfo {author} {\bibfnamefont {A.}~\bibnamefont
  {Carrington}},\ }\href {https://doi.org/DOI: 10.1017/CBO9780511814808} {\emph
  {\bibinfo {title} {Cambridge Molecular Science}}}\ (\bibinfo  {publisher}
  {Cambridge University Press},\ \bibinfo {address} {Cambridge},\ \bibinfo
  {year} {2003})\BibitemShut {NoStop}%
\bibitem [{\citenamefont {Atkins}\ and\ \citenamefont
  {de~Paula}(2014)}]{AtkinsPhysicalChemistry}%
  \BibitemOpen
  \bibfield  {author} {\bibinfo {author} {\bibfnamefont {P.}~\bibnamefont
  {Atkins}}\ and\ \bibinfo {author} {\bibfnamefont {J.}~\bibnamefont
  {de~Paula}},\ }\href@noop {} {\emph {\bibinfo {title} {{Atkins' Physical
  Chemistry}}}}\ (\bibinfo  {publisher} {Oxford University Press},\ \bibinfo
  {year} {2014})\BibitemShut {NoStop}%
\bibitem [{\citenamefont {Satterthwaite}\ \emph {et~al.}(2022)\citenamefont
  {Satterthwaite}, \citenamefont {Koumarianou}, \citenamefont {Sorensen},\ and\
  \citenamefont {Patterson}}]{Satterthwaite2022}%
  \BibitemOpen
  \bibfield  {author} {\bibinfo {author} {\bibfnamefont {L.}~\bibnamefont
  {Satterthwaite}}, \bibinfo {author} {\bibfnamefont {G.}~\bibnamefont
  {Koumarianou}}, \bibinfo {author} {\bibfnamefont {D.}~\bibnamefont
  {Sorensen}},\ and\ \bibinfo {author} {\bibfnamefont {D.}~\bibnamefont
  {Patterson}},\ }\bibfield  {title} {\bibinfo {title} {{Sub-Hz Differential
  Rotational Spectroscopy of Enantiomers}},\ }\bibfield  {journal} {\bibinfo
  {journal} {Symmetry}\ }\textbf {\bibinfo {volume} {14}},\ \href
  {https://doi.org/10.3390/sym14010028} {10.3390/sym14010028} (\bibinfo {year}
  {2022})\BibitemShut {NoStop}%
\bibitem [{\citenamefont {Aiello}\ \emph {et~al.}(2022)\citenamefont {Aiello},
  \citenamefont {Di~Sarno}, \citenamefont {Delli~Santi}, \citenamefont
  {De~Rosa}, \citenamefont {Ricciardi}, \citenamefont {De~Natale},
  \citenamefont {Santamaria}, \citenamefont {Giusfredi},\ and\ \citenamefont
  {Maddaloni}}]{Aiello2022}%
  \BibitemOpen
  \bibfield  {author} {\bibinfo {author} {\bibfnamefont {R.}~\bibnamefont
  {Aiello}}, \bibinfo {author} {\bibfnamefont {V.}~\bibnamefont {Di~Sarno}},
  \bibinfo {author} {\bibfnamefont {M.~G.}\ \bibnamefont {Delli~Santi}},
  \bibinfo {author} {\bibfnamefont {M.}~\bibnamefont {De~Rosa}}, \bibinfo
  {author} {\bibfnamefont {I.}~\bibnamefont {Ricciardi}}, \bibinfo {author}
  {\bibfnamefont {P.}~\bibnamefont {De~Natale}}, \bibinfo {author}
  {\bibfnamefont {L.}~\bibnamefont {Santamaria}}, \bibinfo {author}
  {\bibfnamefont {G.}~\bibnamefont {Giusfredi}},\ and\ \bibinfo {author}
  {\bibfnamefont {P.}~\bibnamefont {Maddaloni}},\ }\bibfield  {title} {\bibinfo
  {title} {{Absolute frequency metrology of buffer-gas-cooled molecular spectra
  at 1 kHz accuracy level}},\ }\href
  {https://doi.org/10.1038/s41467-022-34758-9} {\bibfield  {journal} {\bibinfo
  {journal} {Nature Communications}\ }\textbf {\bibinfo {volume} {13}},\
  \bibinfo {pages} {7016} (\bibinfo {year} {2022})}\BibitemShut {NoStop}%
\bibitem [{\citenamefont {Alighanbari}\ \emph {et~al.}(2020)\citenamefont
  {Alighanbari}, \citenamefont {Giri}, \citenamefont {Constantin},
  \citenamefont {Korobov},\ and\ \citenamefont {Schiller}}]{Alighanbari2020}%
  \BibitemOpen
  \bibfield  {author} {\bibinfo {author} {\bibfnamefont {S.}~\bibnamefont
  {Alighanbari}}, \bibinfo {author} {\bibfnamefont {G.~S.}\ \bibnamefont
  {Giri}}, \bibinfo {author} {\bibfnamefont {F.~L.}\ \bibnamefont
  {Constantin}}, \bibinfo {author} {\bibfnamefont {V.~I.}\ \bibnamefont
  {Korobov}},\ and\ \bibinfo {author} {\bibfnamefont {S.}~\bibnamefont
  {Schiller}},\ }\bibfield  {title} {\bibinfo {title} {{Precise test of quantum
  electrodynamics and determination of fundamental constants with HD+ ions}},\
  }\href {https://doi.org/10.1038/s41586-020-2261-5} {\bibfield  {journal}
  {\bibinfo  {journal} {Nature}\ }\textbf {\bibinfo {volume} {581}},\ \bibinfo
  {pages} {152} (\bibinfo {year} {2020})}\BibitemShut {NoStop}%
\bibitem [{\citenamefont {Patra}\ \emph {et~al.}(2020)\citenamefont {Patra},
  \citenamefont {Germann}, \citenamefont {Karr}, \citenamefont {Haidar},
  \citenamefont {Hilico}, \citenamefont {Korobov}, \citenamefont {Cozijn},
  \citenamefont {Eikema}, \citenamefont {Ubachs},\ and\ \citenamefont
  {Koelemeij}}]{Patra2020}%
  \BibitemOpen
  \bibfield  {author} {\bibinfo {author} {\bibfnamefont {S.}~\bibnamefont
  {Patra}}, \bibinfo {author} {\bibfnamefont {M.}~\bibnamefont {Germann}},
  \bibinfo {author} {\bibfnamefont {J.~P.}\ \bibnamefont {Karr}}, \bibinfo
  {author} {\bibfnamefont {M.}~\bibnamefont {Haidar}}, \bibinfo {author}
  {\bibfnamefont {L.}~\bibnamefont {Hilico}}, \bibinfo {author} {\bibfnamefont
  {V.~I.}\ \bibnamefont {Korobov}}, \bibinfo {author} {\bibfnamefont {F.~M.}\
  \bibnamefont {Cozijn}}, \bibinfo {author} {\bibfnamefont {K.~S.}\
  \bibnamefont {Eikema}}, \bibinfo {author} {\bibfnamefont {W.}~\bibnamefont
  {Ubachs}},\ and\ \bibinfo {author} {\bibfnamefont {J.~C.}\ \bibnamefont
  {Koelemeij}},\ }\bibfield  {title} {\bibinfo {title} {{Proton-electron mass
  ratio from laser spectroscopy of HD+ at the part-per-trillion level}},\
  }\href {https://doi.org/10.1126/SCIENCE.ABA0453} {\bibfield  {journal}
  {\bibinfo  {journal} {Science}\ }\textbf {\bibinfo {volume} {369}},\ \bibinfo
  {pages} {1238} (\bibinfo {year} {2020})}\BibitemShut {NoStop}%
\bibitem [{\citenamefont {Engfer}\ \emph {et~al.}(1974)\citenamefont {Engfer},
  \citenamefont {Schneuwly}, \citenamefont {Vuilleumier}, \citenamefont
  {Walter},\ and\ \citenamefont {Zehnder}}]{Engfer1974}%
  \BibitemOpen
  \bibfield  {author} {\bibinfo {author} {\bibfnamefont {R.}~\bibnamefont
  {Engfer}}, \bibinfo {author} {\bibfnamefont {H.}~\bibnamefont {Schneuwly}},
  \bibinfo {author} {\bibfnamefont {J.~L.}\ \bibnamefont {Vuilleumier}},
  \bibinfo {author} {\bibfnamefont {H.~K.}\ \bibnamefont {Walter}},\ and\
  \bibinfo {author} {\bibfnamefont {A.}~\bibnamefont {Zehnder}},\ }\bibfield
  {title} {\bibinfo {title} {{Charge-distribution parameters, isotope shifts,
  isomer shifts, and magnetic hyperfine constants from muonic atoms}},\ }\href
  {https://doi.org/https://doi.org/10.1016/S0092-640X(74)80003-3} {\bibfield
  {journal} {\bibinfo  {journal} {Atomic Data and Nuclear Data Tables}\
  }\textbf {\bibinfo {volume} {14}},\ \bibinfo {pages} {509} (\bibinfo {year}
  {1974})}\BibitemShut {NoStop}%
\bibitem [{\citenamefont {Hofstadter}(1956)}]{Hofstadter1956}%
  \BibitemOpen
  \bibfield  {author} {\bibinfo {author} {\bibfnamefont {R.}~\bibnamefont
  {Hofstadter}},\ }\bibfield  {title} {\bibinfo {title} {{Electron Scattering
  and Nuclear Structure}},\ }\href {https://doi.org/10.1103/RevModPhys.28.214}
  {\bibfield  {journal} {\bibinfo  {journal} {Reviews of Modern Physics}\
  }\textbf {\bibinfo {volume} {28}},\ \bibinfo {pages} {214} (\bibinfo {year}
  {1956})}\BibitemShut {NoStop}%
\bibitem [{\citenamefont {De~Vries}\ \emph {et~al.}(1987)\citenamefont
  {De~Vries}, \citenamefont {De~Jager},\ and\ \citenamefont
  {De~Vries}}]{deVries1987}%
  \BibitemOpen
  \bibfield  {author} {\bibinfo {author} {\bibfnamefont {H.}~\bibnamefont
  {De~Vries}}, \bibinfo {author} {\bibfnamefont {C.~W.}\ \bibnamefont
  {De~Jager}},\ and\ \bibinfo {author} {\bibfnamefont {C.}~\bibnamefont
  {De~Vries}},\ }\bibfield  {title} {\bibinfo {title} {{Nuclear
  charge-density-distribution parameters from elastic electron scattering}},\
  }\href {https://doi.org/https://doi.org/10.1016/0092-640X(87)90013-1}
  {\bibfield  {journal} {\bibinfo  {journal} {Atomic Data and Nuclear Data
  Tables}\ }\textbf {\bibinfo {volume} {36}},\ \bibinfo {pages} {495} (\bibinfo
  {year} {1987})}\BibitemShut {NoStop}%
\bibitem [{\citenamefont {Angeli}\ and\ \citenamefont
  {Marinova}(2013)}]{Angeli2013}%
  \BibitemOpen
  \bibfield  {author} {\bibinfo {author} {\bibfnamefont {I.}~\bibnamefont
  {Angeli}}\ and\ \bibinfo {author} {\bibfnamefont {K.~P.}\ \bibnamefont
  {Marinova}},\ }\bibfield  {title} {\bibinfo {title} {{Table of experimental
  nuclear ground state charge radii: An update}},\ }\href
  {https://doi.org/https://doi.org/10.1016/j.adt.2011.12.006} {\bibfield
  {journal} {\bibinfo  {journal} {Atomic Data and Nuclear Data Tables}\
  }\textbf {\bibinfo {volume} {99}},\ \bibinfo {pages} {69} (\bibinfo {year}
  {2013})}\BibitemShut {NoStop}%
\bibitem [{\citenamefont {King}(2013)}]{King2013}%
  \BibitemOpen
  \bibfield  {author} {\bibinfo {author} {\bibfnamefont {W.~H.}\ \bibnamefont
  {King}},\ }\href@noop {} {\emph {\bibinfo {title} {{Isotope shifts in atomic
  spectra}}}}\ (\bibinfo  {publisher} {Springer Science {\&} Business Media},\
  \bibinfo {year} {2013})\BibitemShut {NoStop}%
\bibitem [{\citenamefont {Yang}\ \emph {et~al.}(2023)\citenamefont {Yang},
  \citenamefont {Wang}, \citenamefont {Wilkins},\ and\ \citenamefont
  {Garcia~Ruiz}}]{Yang2023}%
  \BibitemOpen
  \bibfield  {author} {\bibinfo {author} {\bibfnamefont {X.~F.}\ \bibnamefont
  {Yang}}, \bibinfo {author} {\bibfnamefont {S.~J.}\ \bibnamefont {Wang}},
  \bibinfo {author} {\bibfnamefont {S.~G.}\ \bibnamefont {Wilkins}},\ and\
  \bibinfo {author} {\bibfnamefont {R.~F.}\ \bibnamefont {Garcia~Ruiz}},\
  }\bibfield  {title} {\bibinfo {title} {{Laser spectroscopy for the study of
  exotic nuclei}},\ }\href
  {https://doi.org/https://doi.org/10.1016/j.ppnp.2022.104005} {\bibfield
  {journal} {\bibinfo  {journal} {Progress in Particle and Nuclear Physics}\
  }\textbf {\bibinfo {volume} {129}},\ \bibinfo {pages} {104005} (\bibinfo
  {year} {2023})}\BibitemShut {NoStop}%
\bibitem [{\citenamefont {Garcia~Ruiz}\ \emph {et~al.}(2016)\citenamefont
  {Garcia~Ruiz}, \citenamefont {Bissell}, \citenamefont {Blaum}, \citenamefont
  {Ekstr{\"{o}}m}, \citenamefont {Fr{\"{o}}mmgen}, \citenamefont {Hagen},
  \citenamefont {Hammen}, \citenamefont {Hebeler}, \citenamefont {Holt},
  \citenamefont {Jansen}, \citenamefont {Kowalska}, \citenamefont {Kreim},
  \citenamefont {Nazarewicz}, \citenamefont {Neugart}, \citenamefont {Neyens},
  \citenamefont {N{\"{o}}rtersh{\"{a}}user}, \citenamefont {Papenbrock},
  \citenamefont {Papuga}, \citenamefont {Schwenk}, \citenamefont {Simonis},
  \citenamefont {Wendt},\ and\ \citenamefont {Yordanov}}]{GarciaRuiz2016}%
  \BibitemOpen
  \bibfield  {author} {\bibinfo {author} {\bibfnamefont {R.~F.}\ \bibnamefont
  {Garcia~Ruiz}}, \bibinfo {author} {\bibfnamefont {M.~L.}\ \bibnamefont
  {Bissell}}, \bibinfo {author} {\bibfnamefont {K.}~\bibnamefont {Blaum}},
  \bibinfo {author} {\bibfnamefont {A.}~\bibnamefont {Ekstr{\"{o}}m}}, \bibinfo
  {author} {\bibfnamefont {N.}~\bibnamefont {Fr{\"{o}}mmgen}}, \bibinfo
  {author} {\bibfnamefont {G.}~\bibnamefont {Hagen}}, \bibinfo {author}
  {\bibfnamefont {M.}~\bibnamefont {Hammen}}, \bibinfo {author} {\bibfnamefont
  {K.}~\bibnamefont {Hebeler}}, \bibinfo {author} {\bibfnamefont {J.~D.}\
  \bibnamefont {Holt}}, \bibinfo {author} {\bibfnamefont {G.~R.}\ \bibnamefont
  {Jansen}}, \bibinfo {author} {\bibfnamefont {M.}~\bibnamefont {Kowalska}},
  \bibinfo {author} {\bibfnamefont {K.}~\bibnamefont {Kreim}}, \bibinfo
  {author} {\bibfnamefont {W.}~\bibnamefont {Nazarewicz}}, \bibinfo {author}
  {\bibfnamefont {R.}~\bibnamefont {Neugart}}, \bibinfo {author} {\bibfnamefont
  {G.}~\bibnamefont {Neyens}}, \bibinfo {author} {\bibfnamefont
  {W.}~\bibnamefont {N{\"{o}}rtersh{\"{a}}user}}, \bibinfo {author}
  {\bibfnamefont {T.}~\bibnamefont {Papenbrock}}, \bibinfo {author}
  {\bibfnamefont {J.}~\bibnamefont {Papuga}}, \bibinfo {author} {\bibfnamefont
  {A.}~\bibnamefont {Schwenk}}, \bibinfo {author} {\bibfnamefont
  {J.}~\bibnamefont {Simonis}}, \bibinfo {author} {\bibfnamefont {K.~A.}\
  \bibnamefont {Wendt}},\ and\ \bibinfo {author} {\bibfnamefont {D.~T.}\
  \bibnamefont {Yordanov}},\ }\bibfield  {title} {\bibinfo {title}
  {{Unexpectedly large charge radii of neutron-rich calcium isotopes}},\ }\href
  {https://doi.org/10.1038/nphys3645} {\bibfield  {journal} {\bibinfo
  {journal} {Nature Physics}\ }\textbf {\bibinfo {volume} {12}},\ \bibinfo
  {pages} {594} (\bibinfo {year} {2016})}\BibitemShut {NoStop}%
\bibitem [{\citenamefont {Koszor{\'{u}}s}\ \emph {et~al.}(2021)\citenamefont
  {Koszor{\'{u}}s}, \citenamefont {Yang}, \citenamefont {Jiang}, \citenamefont
  {Novario}, \citenamefont {Bai}, \citenamefont {Billowes}, \citenamefont
  {Binnersley}, \citenamefont {Bissell}, \citenamefont {Cocolios},
  \citenamefont {Cooper}, \citenamefont {de~Groote}, \citenamefont
  {Ekstr{\"{o}}m}, \citenamefont {Flanagan}, \citenamefont {Forss{\'{e}}n},
  \citenamefont {Franchoo}, \citenamefont {Ruiz}, \citenamefont {Gustafsson},
  \citenamefont {Hagen}, \citenamefont {Jansen}, \citenamefont
  {Kanellakopoulos}, \citenamefont {Kortelainen}, \citenamefont {Nazarewicz},
  \citenamefont {Neyens}, \citenamefont {Papenbrock}, \citenamefont {Reinhard},
  \citenamefont {Ricketts}, \citenamefont {Sahoo}, \citenamefont {Vernon},\
  and\ \citenamefont {Wilkins}}]{Koszorus2021}%
  \BibitemOpen
  \bibfield  {author} {\bibinfo {author} {\bibfnamefont {A.}~\bibnamefont
  {Koszor{\'{u}}s}}, \bibinfo {author} {\bibfnamefont {X.~F.}\ \bibnamefont
  {Yang}}, \bibinfo {author} {\bibfnamefont {W.~G.}\ \bibnamefont {Jiang}},
  \bibinfo {author} {\bibfnamefont {S.~J.}\ \bibnamefont {Novario}}, \bibinfo
  {author} {\bibfnamefont {S.~W.}\ \bibnamefont {Bai}}, \bibinfo {author}
  {\bibfnamefont {J.}~\bibnamefont {Billowes}}, \bibinfo {author}
  {\bibfnamefont {C.~L.}\ \bibnamefont {Binnersley}}, \bibinfo {author}
  {\bibfnamefont {M.~L.}\ \bibnamefont {Bissell}}, \bibinfo {author}
  {\bibfnamefont {T.~E.}\ \bibnamefont {Cocolios}}, \bibinfo {author}
  {\bibfnamefont {B.~S.}\ \bibnamefont {Cooper}}, \bibinfo {author}
  {\bibfnamefont {R.~P.}\ \bibnamefont {de~Groote}}, \bibinfo {author}
  {\bibfnamefont {A.}~\bibnamefont {Ekstr{\"{o}}m}}, \bibinfo {author}
  {\bibfnamefont {K.~T.}\ \bibnamefont {Flanagan}}, \bibinfo {author}
  {\bibfnamefont {C.}~\bibnamefont {Forss{\'{e}}n}}, \bibinfo {author}
  {\bibfnamefont {S.}~\bibnamefont {Franchoo}}, \bibinfo {author}
  {\bibfnamefont {R.~F.~G.}\ \bibnamefont {Ruiz}}, \bibinfo {author}
  {\bibfnamefont {F.~P.}\ \bibnamefont {Gustafsson}}, \bibinfo {author}
  {\bibfnamefont {G.}~\bibnamefont {Hagen}}, \bibinfo {author} {\bibfnamefont
  {G.~R.}\ \bibnamefont {Jansen}}, \bibinfo {author} {\bibfnamefont
  {A.}~\bibnamefont {Kanellakopoulos}}, \bibinfo {author} {\bibfnamefont
  {M.}~\bibnamefont {Kortelainen}}, \bibinfo {author} {\bibfnamefont
  {W.}~\bibnamefont {Nazarewicz}}, \bibinfo {author} {\bibfnamefont
  {G.}~\bibnamefont {Neyens}}, \bibinfo {author} {\bibfnamefont
  {T.}~\bibnamefont {Papenbrock}}, \bibinfo {author} {\bibfnamefont {P.-G.}\
  \bibnamefont {Reinhard}}, \bibinfo {author} {\bibfnamefont {C.~M.}\
  \bibnamefont {Ricketts}}, \bibinfo {author} {\bibfnamefont {B.~K.}\
  \bibnamefont {Sahoo}}, \bibinfo {author} {\bibfnamefont {A.~R.}\ \bibnamefont
  {Vernon}},\ and\ \bibinfo {author} {\bibfnamefont {S.~G.}\ \bibnamefont
  {Wilkins}},\ }\bibfield  {title} {\bibinfo {title} {{Charge radii of exotic
  potassium isotopes challenge nuclear theory and the magic character of N =
  32}},\ }\bibfield  {journal} {\bibinfo  {journal} {Nature Physics}\ }\href
  {https://doi.org/10.1038/s41567-020-01136-5} {10.1038/s41567-020-01136-5}
  (\bibinfo {year} {2021})\BibitemShut {NoStop}%
\bibitem [{\citenamefont {de~Groote}\ \emph {et~al.}(2020)\citenamefont
  {de~Groote}, \citenamefont {Billowes}, \citenamefont {Binnersley},
  \citenamefont {Bissell}, \citenamefont {Cocolios}, \citenamefont
  {Day~Goodacre}, \citenamefont {Farooq-Smith}, \citenamefont {Fedorov},
  \citenamefont {Flanagan}, \citenamefont {Franchoo}, \citenamefont
  {Garcia~Ruiz}, \citenamefont {Gins}, \citenamefont {Holt}, \citenamefont
  {{Koszor{\'{u}}s}}, \citenamefont {Lynch}, \citenamefont {Miyagi},
  \citenamefont {Nazarewicz}, \citenamefont {Neyens}, \citenamefont {Reinhard},
  \citenamefont {Rothe}, \citenamefont {Stroke}, \citenamefont {Vernon},
  \citenamefont {Wendt}, \citenamefont {Wilkins}, \citenamefont {Xu},\ and\
  \citenamefont {Yang}}]{deGroote2020b}%
  \BibitemOpen
  \bibfield  {author} {\bibinfo {author} {\bibfnamefont {R.~P.}\ \bibnamefont
  {de~Groote}}, \bibinfo {author} {\bibfnamefont {J.}~\bibnamefont {Billowes}},
  \bibinfo {author} {\bibfnamefont {C.~L.}\ \bibnamefont {Binnersley}},
  \bibinfo {author} {\bibfnamefont {M.~L.}\ \bibnamefont {Bissell}}, \bibinfo
  {author} {\bibfnamefont {T.~E.}\ \bibnamefont {Cocolios}}, \bibinfo {author}
  {\bibfnamefont {T.}~\bibnamefont {Day~Goodacre}}, \bibinfo {author}
  {\bibfnamefont {G.~J.}\ \bibnamefont {Farooq-Smith}}, \bibinfo {author}
  {\bibfnamefont {D.~V.}\ \bibnamefont {Fedorov}}, \bibinfo {author}
  {\bibfnamefont {K.~T.}\ \bibnamefont {Flanagan}}, \bibinfo {author}
  {\bibfnamefont {S.}~\bibnamefont {Franchoo}}, \bibinfo {author}
  {\bibfnamefont {R.~F.}\ \bibnamefont {Garcia~Ruiz}}, \bibinfo {author}
  {\bibfnamefont {W.}~\bibnamefont {Gins}}, \bibinfo {author} {\bibfnamefont
  {J.~D.}\ \bibnamefont {Holt}}, \bibinfo {author} {\bibnamefont
  {{Koszor{\'{u}}s}}}, \bibinfo {author} {\bibfnamefont {K.~M.}\ \bibnamefont
  {Lynch}}, \bibinfo {author} {\bibfnamefont {T.}~\bibnamefont {Miyagi}},
  \bibinfo {author} {\bibfnamefont {W.}~\bibnamefont {Nazarewicz}}, \bibinfo
  {author} {\bibfnamefont {G.}~\bibnamefont {Neyens}}, \bibinfo {author}
  {\bibfnamefont {P.~G.}\ \bibnamefont {Reinhard}}, \bibinfo {author}
  {\bibfnamefont {S.}~\bibnamefont {Rothe}}, \bibinfo {author} {\bibfnamefont
  {H.~H.}\ \bibnamefont {Stroke}}, \bibinfo {author} {\bibfnamefont {A.~R.}\
  \bibnamefont {Vernon}}, \bibinfo {author} {\bibfnamefont {K.~D.}\
  \bibnamefont {Wendt}}, \bibinfo {author} {\bibfnamefont {S.~G.}\ \bibnamefont
  {Wilkins}}, \bibinfo {author} {\bibfnamefont {Z.~Y.}\ \bibnamefont {Xu}},\
  and\ \bibinfo {author} {\bibfnamefont {X.~F.}\ \bibnamefont {Yang}},\
  }\bibfield  {title} {\bibinfo {title} {{Measurement and microscopic
  description of odd–even staggering of charge radii of exotic copper
  isotopes}},\ }\href {https://doi.org/10.1038/s41567-020-0868-y} {\bibfield
  {journal} {\bibinfo  {journal} {Nature Physics}\ }\textbf {\bibinfo {volume}
  {16}},\ \bibinfo {pages} {620} (\bibinfo {year} {2020})}\BibitemShut
  {NoStop}%
\bibitem [{\citenamefont {Verstraelen}\ \emph {et~al.}(2019)\citenamefont
  {Verstraelen}, \citenamefont {Teigelh{\"{o}}fer}, \citenamefont {Ryssens},
  \citenamefont {Ames}, \citenamefont {Barzakh}, \citenamefont {Bender},
  \citenamefont {Ferrer}, \citenamefont {Goriely}, \citenamefont {Heenen},
  \citenamefont {Huyse}, \citenamefont {Kunz}, \citenamefont {Lassen},
  \citenamefont {Manea}, \citenamefont {Raeder},\ and\ \citenamefont
  {Van~Duppen}}]{Verstraelen2019a}%
  \BibitemOpen
  \bibfield  {author} {\bibinfo {author} {\bibfnamefont {E.}~\bibnamefont
  {Verstraelen}}, \bibinfo {author} {\bibfnamefont {A.}~\bibnamefont
  {Teigelh{\"{o}}fer}}, \bibinfo {author} {\bibfnamefont {W.}~\bibnamefont
  {Ryssens}}, \bibinfo {author} {\bibfnamefont {F.}~\bibnamefont {Ames}},
  \bibinfo {author} {\bibfnamefont {A.}~\bibnamefont {Barzakh}}, \bibinfo
  {author} {\bibfnamefont {M.}~\bibnamefont {Bender}}, \bibinfo {author}
  {\bibfnamefont {R.}~\bibnamefont {Ferrer}}, \bibinfo {author} {\bibfnamefont
  {S.}~\bibnamefont {Goriely}}, \bibinfo {author} {\bibfnamefont {P.~H.}\
  \bibnamefont {Heenen}}, \bibinfo {author} {\bibfnamefont {M.}~\bibnamefont
  {Huyse}}, \bibinfo {author} {\bibfnamefont {P.}~\bibnamefont {Kunz}},
  \bibinfo {author} {\bibfnamefont {J.}~\bibnamefont {Lassen}}, \bibinfo
  {author} {\bibfnamefont {V.}~\bibnamefont {Manea}}, \bibinfo {author}
  {\bibfnamefont {S.}~\bibnamefont {Raeder}},\ and\ \bibinfo {author}
  {\bibfnamefont {P.}~\bibnamefont {Van~Duppen}},\ }\bibfield  {title}
  {\bibinfo {title} {{Search for octupole-deformed actinium isotopes using
  resonance ionization spectroscopy}},\ }\href
  {https://doi.org/10.1103/PhysRevC.100.044321} {\bibfield  {journal} {\bibinfo
   {journal} {Physical Review C}\ }\textbf {\bibinfo {volume} {100}},\ \bibinfo
  {pages} {44321} (\bibinfo {year} {2019})}\BibitemShut {NoStop}%
\bibitem [{\citenamefont {Miller}\ \emph {et~al.}(2019)\citenamefont {Miller},
  \citenamefont {Minamisono}, \citenamefont {Klose}, \citenamefont {Garand},
  \citenamefont {Kujawa}, \citenamefont {Lantis}, \citenamefont {Liu},
  \citenamefont {Maa{\ss}}, \citenamefont {Mantica}, \citenamefont
  {Nazarewicz}, \citenamefont {N{\"{o}}rtersh{\"{a}}user}, \citenamefont
  {Pineda}, \citenamefont {Reinhard}, \citenamefont {Rossi}, \citenamefont
  {Sommer}, \citenamefont {Sumithrarachchi}, \citenamefont
  {Teigelh{\"{o}}fer},\ and\ \citenamefont {Watkins}}]{Miller2019}%
  \BibitemOpen
  \bibfield  {author} {\bibinfo {author} {\bibfnamefont {A.~J.}\ \bibnamefont
  {Miller}}, \bibinfo {author} {\bibfnamefont {K.}~\bibnamefont {Minamisono}},
  \bibinfo {author} {\bibfnamefont {A.}~\bibnamefont {Klose}}, \bibinfo
  {author} {\bibfnamefont {D.}~\bibnamefont {Garand}}, \bibinfo {author}
  {\bibfnamefont {C.}~\bibnamefont {Kujawa}}, \bibinfo {author} {\bibfnamefont
  {J.~D.}\ \bibnamefont {Lantis}}, \bibinfo {author} {\bibfnamefont
  {Y.}~\bibnamefont {Liu}}, \bibinfo {author} {\bibfnamefont {B.}~\bibnamefont
  {Maa{\ss}}}, \bibinfo {author} {\bibfnamefont {P.~F.}\ \bibnamefont
  {Mantica}}, \bibinfo {author} {\bibfnamefont {W.}~\bibnamefont {Nazarewicz}},
  \bibinfo {author} {\bibfnamefont {W.}~\bibnamefont
  {N{\"{o}}rtersh{\"{a}}user}}, \bibinfo {author} {\bibfnamefont {S.~V.}\
  \bibnamefont {Pineda}}, \bibinfo {author} {\bibfnamefont {P.-G.}\
  \bibnamefont {Reinhard}}, \bibinfo {author} {\bibfnamefont {D.~M.}\
  \bibnamefont {Rossi}}, \bibinfo {author} {\bibfnamefont {F.}~\bibnamefont
  {Sommer}}, \bibinfo {author} {\bibfnamefont {C.}~\bibnamefont
  {Sumithrarachchi}}, \bibinfo {author} {\bibfnamefont {A.}~\bibnamefont
  {Teigelh{\"{o}}fer}},\ and\ \bibinfo {author} {\bibfnamefont
  {J.}~\bibnamefont {Watkins}},\ }\bibfield  {title} {\bibinfo {title} {{Proton
  superfluidity and charge radii in proton-rich calcium isotopes}},\ }\href
  {https://doi.org/10.1038/s41567-019-0416-9} {\bibfield  {journal} {\bibinfo
  {journal} {Nature Physics}\ }\textbf {\bibinfo {volume} {15}},\ \bibinfo
  {pages} {432} (\bibinfo {year} {2019})}\BibitemShut {NoStop}%
\bibitem [{\citenamefont {Adamczak}\ \emph {et~al.}(2018)\citenamefont
  {Adamczak}, \citenamefont {Antognini}, \citenamefont {Berger}, \citenamefont
  {Cocolios}, \citenamefont {Dressler}, \citenamefont {Eggenberger},
  \citenamefont {Eichler}, \citenamefont {Indelicato}, \citenamefont
  {Jungmann}, \citenamefont {Kirch}, \citenamefont {Knecht}, \citenamefont
  {Papa}, \citenamefont {Pohl}, \citenamefont {Pospelov}, \citenamefont
  {Rapisarda}, \citenamefont {Reiter}, \citenamefont {Ritjoho}, \citenamefont
  {Roccia}, \citenamefont {Severijns}, \citenamefont {Skawran}, \citenamefont
  {Wauters},\ and\ \citenamefont {Willmann}}]{Adamczak2018}%
  \BibitemOpen
  \bibfield  {author} {\bibinfo {author} {\bibfnamefont {A.}~\bibnamefont
  {Adamczak}}, \bibinfo {author} {\bibfnamefont {A.}~\bibnamefont {Antognini}},
  \bibinfo {author} {\bibfnamefont {N.}~\bibnamefont {Berger}}, \bibinfo
  {author} {\bibfnamefont {T.~E.}\ \bibnamefont {Cocolios}}, \bibinfo {author}
  {\bibfnamefont {R.}~\bibnamefont {Dressler}}, \bibinfo {author}
  {\bibfnamefont {A.}~\bibnamefont {Eggenberger}}, \bibinfo {author}
  {\bibfnamefont {R.}~\bibnamefont {Eichler}}, \bibinfo {author} {\bibfnamefont
  {P.}~\bibnamefont {Indelicato}}, \bibinfo {author} {\bibfnamefont
  {K.}~\bibnamefont {Jungmann}}, \bibinfo {author} {\bibfnamefont
  {K.}~\bibnamefont {Kirch}}, \bibinfo {author} {\bibfnamefont
  {A.}~\bibnamefont {Knecht}}, \bibinfo {author} {\bibfnamefont
  {A.}~\bibnamefont {Papa}}, \bibinfo {author} {\bibfnamefont {R.}~\bibnamefont
  {Pohl}}, \bibinfo {author} {\bibfnamefont {M.}~\bibnamefont {Pospelov}},
  \bibinfo {author} {\bibfnamefont {E.}~\bibnamefont {Rapisarda}}, \bibinfo
  {author} {\bibfnamefont {P.}~\bibnamefont {Reiter}}, \bibinfo {author}
  {\bibfnamefont {N.}~\bibnamefont {Ritjoho}}, \bibinfo {author} {\bibfnamefont
  {S.}~\bibnamefont {Roccia}}, \bibinfo {author} {\bibfnamefont
  {N.}~\bibnamefont {Severijns}}, \bibinfo {author} {\bibfnamefont
  {A.}~\bibnamefont {Skawran}}, \bibinfo {author} {\bibfnamefont
  {F.}~\bibnamefont {Wauters}},\ and\ \bibinfo {author} {\bibfnamefont
  {L.}~\bibnamefont {Willmann}},\ }\bibfield  {title} {\bibinfo {title}
  {{Nuclear structure with radioactive muonic atoms}},\ }\href
  {https://doi.org/10.1051/epjconf/201819304014} {\bibfield  {journal}
  {\bibinfo  {journal} {EPJ Web Conf.}\ }\textbf {\bibinfo {volume} {193}}
  (\bibinfo {year} {2018})}\BibitemShut {NoStop}%
\bibitem [{\citenamefont {Tsukada}\ \emph {et~al.}(2023)\citenamefont
  {Tsukada}, \citenamefont {Abe}, \citenamefont {Enokizono}, \citenamefont
  {Goke}, \citenamefont {Hara}, \citenamefont {Honda}, \citenamefont {Hori},
  \citenamefont {Ichikawa}, \citenamefont {Ito}, \citenamefont {Kurita},
  \citenamefont {Legris}, \citenamefont {Maehara}, \citenamefont {Ohnishi},
  \citenamefont {Ogawara}, \citenamefont {Suda}, \citenamefont {Tamae},
  \citenamefont {Wakasugi}, \citenamefont {Watanabe},\ and\ \citenamefont
  {Wauke}}]{SCRIT2023}%
  \BibitemOpen
  \bibfield  {author} {\bibinfo {author} {\bibfnamefont {K.}~\bibnamefont
  {Tsukada}}, \bibinfo {author} {\bibfnamefont {Y.}~\bibnamefont {Abe}},
  \bibinfo {author} {\bibfnamefont {A.}~\bibnamefont {Enokizono}}, \bibinfo
  {author} {\bibfnamefont {T.}~\bibnamefont {Goke}}, \bibinfo {author}
  {\bibfnamefont {M.}~\bibnamefont {Hara}}, \bibinfo {author} {\bibfnamefont
  {Y.}~\bibnamefont {Honda}}, \bibinfo {author} {\bibfnamefont
  {T.}~\bibnamefont {Hori}}, \bibinfo {author} {\bibfnamefont {S.}~\bibnamefont
  {Ichikawa}}, \bibinfo {author} {\bibfnamefont {Y.}~\bibnamefont {Ito}},
  \bibinfo {author} {\bibfnamefont {K.}~\bibnamefont {Kurita}}, \bibinfo
  {author} {\bibfnamefont {C.}~\bibnamefont {Legris}}, \bibinfo {author}
  {\bibfnamefont {Y.}~\bibnamefont {Maehara}}, \bibinfo {author} {\bibfnamefont
  {T.}~\bibnamefont {Ohnishi}}, \bibinfo {author} {\bibfnamefont
  {R.}~\bibnamefont {Ogawara}}, \bibinfo {author} {\bibfnamefont
  {T.}~\bibnamefont {Suda}}, \bibinfo {author} {\bibfnamefont {T.}~\bibnamefont
  {Tamae}}, \bibinfo {author} {\bibfnamefont {M.}~\bibnamefont {Wakasugi}},
  \bibinfo {author} {\bibfnamefont {M.}~\bibnamefont {Watanabe}},\ and\
  \bibinfo {author} {\bibfnamefont {H.}~\bibnamefont {Wauke}},\ }\bibfield
  {title} {\bibinfo {title} {{First Observation of Electron Scattering from
  Online-Produced Radioactive Target}},\ }\bibfield  {journal} {\bibinfo
  {journal} {Physical Review Letters}\ }\textbf {\bibinfo {volume} {131}},\
  \href {https://doi.org/10.1103/PhysRevLett.131.092502}
  {10.1103/PhysRevLett.131.092502} (\bibinfo {year} {2023})\BibitemShut
  {NoStop}%
\bibitem [{\citenamefont {Brissaud}\ \emph {et~al.}(1972)\citenamefont
  {Brissaud}, \citenamefont {Le~Bornec}, \citenamefont {Tatischeff},
  \citenamefont {Bimbot}, \citenamefont {Brussel},\ and\ \citenamefont
  {Duhamel}}]{Brissaud1972}%
  \BibitemOpen
  \bibfield  {author} {\bibinfo {author} {\bibfnamefont {I.}~\bibnamefont
  {Brissaud}}, \bibinfo {author} {\bibfnamefont {Y.}~\bibnamefont {Le~Bornec}},
  \bibinfo {author} {\bibfnamefont {B.}~\bibnamefont {Tatischeff}}, \bibinfo
  {author} {\bibfnamefont {L.}~\bibnamefont {Bimbot}}, \bibinfo {author}
  {\bibfnamefont {M.~K.}\ \bibnamefont {Brussel}},\ and\ \bibinfo {author}
  {\bibfnamefont {G.}~\bibnamefont {Duhamel}},\ }\bibfield  {title} {\bibinfo
  {title} {{D{\'{e}}termination du rayon de la distribution de neutrons de
  certains noyaux par l'{\'{e}}tude de la diffusion {\'{e}}lastique de
  particules alpha de 166 mev}},\ }\href
  {https://doi.org/https://doi.org/10.1016/0375-9474(72)90599-4} {\bibfield
  {journal} {\bibinfo  {journal} {Nuclear Physics A}\ }\textbf {\bibinfo
  {volume} {191}},\ \bibinfo {pages} {145} (\bibinfo {year}
  {1972})}\BibitemShut {NoStop}%
\bibitem [{\citenamefont {Alkhazov}\ \emph {et~al.}(1978)\citenamefont
  {Alkhazov}, \citenamefont {Belostotsky},\ and\ \citenamefont
  {Vorobyov}}]{Alkhazov1978}%
  \BibitemOpen
  \bibfield  {author} {\bibinfo {author} {\bibfnamefont {G.~D.}\ \bibnamefont
  {Alkhazov}}, \bibinfo {author} {\bibfnamefont {S.~L.}\ \bibnamefont
  {Belostotsky}},\ and\ \bibinfo {author} {\bibfnamefont {A.~A.}\ \bibnamefont
  {Vorobyov}},\ }\bibfield  {title} {\bibinfo {title} {{Scattering of 1 GeV
  protons on nuclei}},\ }\href
  {https://doi.org/https://doi.org/10.1016/0370-1573(78)90083-2} {\bibfield
  {journal} {\bibinfo  {journal} {Physics Reports}\ }\textbf {\bibinfo {volume}
  {42}},\ \bibinfo {pages} {89} (\bibinfo {year} {1978})}\BibitemShut {NoStop}%
\bibitem [{\citenamefont {Rutherford}(1911)}]{Rutherford1911}%
  \BibitemOpen
  \bibfield  {author} {\bibinfo {author} {\bibfnamefont {E.}~\bibnamefont
  {Rutherford}},\ }\bibfield  {title} {\bibinfo {title} {{LXXIX. The scattering
  of {$\alpha$} and {$\beta$} particles by matter and the structure of the
  atom}},\ }\href {https://doi.org/10.1080/14786440508637080} {\bibfield
  {journal} {\bibinfo  {journal} {The London, Edinburgh, and Dublin
  Philosophical Magazine and Journal of Science}\ }\textbf {\bibinfo {volume}
  {21}},\ \bibinfo {pages} {669} (\bibinfo {year} {1911})}\BibitemShut
  {NoStop}%
\bibitem [{\citenamefont {Batty}\ \emph {et~al.}(1989)\citenamefont {Batty},
  \citenamefont {Friedman}, \citenamefont {Gils},\ and\ \citenamefont
  {Rebel}}]{Batty1989}%
  \BibitemOpen
  \bibfield  {author} {\bibinfo {author} {\bibfnamefont {C.~J.}\ \bibnamefont
  {Batty}}, \bibinfo {author} {\bibfnamefont {E.}~\bibnamefont {Friedman}},
  \bibinfo {author} {\bibfnamefont {H.~J.}\ \bibnamefont {Gils}},\ and\
  \bibinfo {author} {\bibfnamefont {H.}~\bibnamefont {Rebel}},\ }\bibfield
  {title} {\bibinfo {title} {{Experimental Methods for Studying Nuclear Density
  Distributions}},\ }in\ \href {https://doi.org/10.1007/978-1-4613-9907-0{\_}1}
  {\emph {\bibinfo {booktitle} {Advances in Nuclear Physics}}},\ \bibinfo
  {editor} {edited by\ \bibinfo {editor} {\bibfnamefont {J.~W.}\ \bibnamefont
  {Negele}}\ and\ \bibinfo {editor} {\bibfnamefont {E.}~\bibnamefont {Vogt}}}\
  (\bibinfo  {publisher} {Springer US},\ \bibinfo {address} {Boston, MA},\
  \bibinfo {year} {1989})\ pp.\ \bibinfo {pages} {1--188}\BibitemShut {NoStop}%
\bibitem [{\citenamefont {Aumann}\ \emph {et~al.}(2022)\citenamefont {Aumann},
  \citenamefont {Bartmann}, \citenamefont {Boine-Frankenheim}, \citenamefont
  {Bouvard}, \citenamefont {Broche}, \citenamefont {Butin}, \citenamefont
  {Calvet}, \citenamefont {Carbonell}, \citenamefont {Chiggiato}, \citenamefont
  {De~Gersem}, \citenamefont {De~Oliveira}, \citenamefont {Dobers},
  \citenamefont {Ehm}, \citenamefont {Somoza}, \citenamefont {Fischer},
  \citenamefont {Fraser}, \citenamefont {Friedrich}, \citenamefont {Frotscher},
  \citenamefont {Gomez-Ramos}, \citenamefont {Grenard}, \citenamefont {Hobl},
  \citenamefont {Hupin}, \citenamefont {Husson}, \citenamefont {Indelicato},
  \citenamefont {Johnston}, \citenamefont {Klink}, \citenamefont {Kubota},
  \citenamefont {Lazauskas}, \citenamefont {Malbrunot-Ettenauer}, \citenamefont
  {Marsic}, \citenamefont {O~M{\"{u}}ller}, \citenamefont {Naimi},
  \citenamefont {Nakatsuka}, \citenamefont {Necca}, \citenamefont {Neidherr},
  \citenamefont {Neyens}, \citenamefont {Obertelli}, \citenamefont {Ono},
  \citenamefont {Pasinelli}, \citenamefont {Paul}, \citenamefont {Pollacco},
  \citenamefont {Rossi}, \citenamefont {Scheit}, \citenamefont {Schlaich},
  \citenamefont {Schmidt}, \citenamefont {Schweikhard}, \citenamefont {Seki},
  \citenamefont {Sels}, \citenamefont {Siesling}, \citenamefont {Uesaka},
  \citenamefont {Vil{\'{e}}n}, \citenamefont {Wada}, \citenamefont {Wienholtz},
  \citenamefont {Wycech},\ and\ \citenamefont {Zacarias}}]{Aumann2022}%
  \BibitemOpen
  \bibfield  {author} {\bibinfo {author} {\bibfnamefont {T.}~\bibnamefont
  {Aumann}}, \bibinfo {author} {\bibfnamefont {W.}~\bibnamefont {Bartmann}},
  \bibinfo {author} {\bibfnamefont {O.}~\bibnamefont {Boine-Frankenheim}},
  \bibinfo {author} {\bibfnamefont {A.}~\bibnamefont {Bouvard}}, \bibinfo
  {author} {\bibfnamefont {A.}~\bibnamefont {Broche}}, \bibinfo {author}
  {\bibfnamefont {F.}~\bibnamefont {Butin}}, \bibinfo {author} {\bibfnamefont
  {D.}~\bibnamefont {Calvet}}, \bibinfo {author} {\bibfnamefont
  {J.}~\bibnamefont {Carbonell}}, \bibinfo {author} {\bibfnamefont
  {P.}~\bibnamefont {Chiggiato}}, \bibinfo {author} {\bibfnamefont
  {H.}~\bibnamefont {De~Gersem}}, \bibinfo {author} {\bibfnamefont
  {R.}~\bibnamefont {De~Oliveira}}, \bibinfo {author} {\bibfnamefont
  {T.}~\bibnamefont {Dobers}}, \bibinfo {author} {\bibfnamefont
  {F.}~\bibnamefont {Ehm}}, \bibinfo {author} {\bibfnamefont {J.~F.}\
  \bibnamefont {Somoza}}, \bibinfo {author} {\bibfnamefont {J.}~\bibnamefont
  {Fischer}}, \bibinfo {author} {\bibfnamefont {M.}~\bibnamefont {Fraser}},
  \bibinfo {author} {\bibfnamefont {E.}~\bibnamefont {Friedrich}}, \bibinfo
  {author} {\bibfnamefont {A.}~\bibnamefont {Frotscher}}, \bibinfo {author}
  {\bibfnamefont {M.}~\bibnamefont {Gomez-Ramos}}, \bibinfo {author}
  {\bibfnamefont {J.-L.}\ \bibnamefont {Grenard}}, \bibinfo {author}
  {\bibfnamefont {A.}~\bibnamefont {Hobl}}, \bibinfo {author} {\bibfnamefont
  {G.}~\bibnamefont {Hupin}}, \bibinfo {author} {\bibfnamefont
  {A.}~\bibnamefont {Husson}}, \bibinfo {author} {\bibfnamefont
  {P.}~\bibnamefont {Indelicato}}, \bibinfo {author} {\bibfnamefont
  {K.}~\bibnamefont {Johnston}}, \bibinfo {author} {\bibfnamefont
  {C.}~\bibnamefont {Klink}}, \bibinfo {author} {\bibfnamefont
  {Y.}~\bibnamefont {Kubota}}, \bibinfo {author} {\bibfnamefont
  {R.}~\bibnamefont {Lazauskas}}, \bibinfo {author} {\bibfnamefont
  {S.}~\bibnamefont {Malbrunot-Ettenauer}}, \bibinfo {author} {\bibfnamefont
  {N.}~\bibnamefont {Marsic}}, \bibinfo {author} {\bibfnamefont {W.~F.}\
  \bibnamefont {O~M{\"{u}}ller}}, \bibinfo {author} {\bibfnamefont
  {S.}~\bibnamefont {Naimi}}, \bibinfo {author} {\bibfnamefont
  {N.}~\bibnamefont {Nakatsuka}}, \bibinfo {author} {\bibfnamefont
  {R.}~\bibnamefont {Necca}}, \bibinfo {author} {\bibfnamefont
  {D.}~\bibnamefont {Neidherr}}, \bibinfo {author} {\bibfnamefont
  {G.}~\bibnamefont {Neyens}}, \bibinfo {author} {\bibfnamefont
  {A.}~\bibnamefont {Obertelli}}, \bibinfo {author} {\bibfnamefont
  {Y.}~\bibnamefont {Ono}}, \bibinfo {author} {\bibfnamefont {S.}~\bibnamefont
  {Pasinelli}}, \bibinfo {author} {\bibfnamefont {N.}~\bibnamefont {Paul}},
  \bibinfo {author} {\bibfnamefont {E.~C.}\ \bibnamefont {Pollacco}}, \bibinfo
  {author} {\bibfnamefont {D.}~\bibnamefont {Rossi}}, \bibinfo {author}
  {\bibfnamefont {H.}~\bibnamefont {Scheit}}, \bibinfo {author} {\bibfnamefont
  {M.}~\bibnamefont {Schlaich}}, \bibinfo {author} {\bibfnamefont
  {A.}~\bibnamefont {Schmidt}}, \bibinfo {author} {\bibfnamefont
  {L.}~\bibnamefont {Schweikhard}}, \bibinfo {author} {\bibfnamefont
  {R.}~\bibnamefont {Seki}}, \bibinfo {author} {\bibfnamefont {S.}~\bibnamefont
  {Sels}}, \bibinfo {author} {\bibfnamefont {E.}~\bibnamefont {Siesling}},
  \bibinfo {author} {\bibfnamefont {T.}~\bibnamefont {Uesaka}}, \bibinfo
  {author} {\bibfnamefont {M.}~\bibnamefont {Vil{\'{e}}n}}, \bibinfo {author}
  {\bibfnamefont {M.}~\bibnamefont {Wada}}, \bibinfo {author} {\bibfnamefont
  {F.}~\bibnamefont {Wienholtz}}, \bibinfo {author} {\bibfnamefont
  {S.}~\bibnamefont {Wycech}},\ and\ \bibinfo {author} {\bibfnamefont
  {S.}~\bibnamefont {Zacarias}},\ }\bibfield  {title} {\bibinfo {title} {{PUMA,
  antiProton unstable matter annihilation}},\ }\href
  {https://doi.org/10.1140/epja/s10050-022-00713-x} {\bibfield  {journal}
  {\bibinfo  {journal} {The European Physical Journal A}\ }\textbf {\bibinfo
  {volume} {58}},\ \bibinfo {pages} {88} (\bibinfo {year} {2022})}\BibitemShut
  {NoStop}%
\bibitem [{\citenamefont {Thiel}\ \emph {et~al.}(2019)\citenamefont {Thiel},
  \citenamefont {Sfienti}, \citenamefont {Piekarewicz}, \citenamefont
  {Horowitz},\ and\ \citenamefont {Vanderhaeghen}}]{Thiel2019}%
  \BibitemOpen
  \bibfield  {author} {\bibinfo {author} {\bibfnamefont {M.}~\bibnamefont
  {Thiel}}, \bibinfo {author} {\bibfnamefont {C.}~\bibnamefont {Sfienti}},
  \bibinfo {author} {\bibfnamefont {J.}~\bibnamefont {Piekarewicz}}, \bibinfo
  {author} {\bibfnamefont {C.~J.}\ \bibnamefont {Horowitz}},\ and\ \bibinfo
  {author} {\bibfnamefont {M.}~\bibnamefont {Vanderhaeghen}},\ }\bibfield
  {title} {\bibinfo {title} {{Neutron skins of atomic nuclei: per aspera ad
  astra}},\ }\href {https://doi.org/10.1088/1361-6471/ab2c6d} {\bibfield
  {journal} {\bibinfo  {journal} {Journal of Physics G: Nuclear and Particle
  Physics}\ }\textbf {\bibinfo {volume} {46}},\ \bibinfo {pages} {093003}
  (\bibinfo {year} {2019})}\BibitemShut {NoStop}%
\bibitem [{\citenamefont {Adhikari}\ \emph {et~al.}(2021)\citenamefont
  {Adhikari}, \citenamefont {Albataineh}, \citenamefont {Androic},
  \citenamefont {Aniol}, \citenamefont {Armstrong}, \citenamefont {Averett},
  \citenamefont {Ayerbe~Gayoso}, \citenamefont {Barcus}, \citenamefont
  {Bellini}, \citenamefont {Beminiwattha}, \citenamefont {Benesch},
  \citenamefont {Bhatt}, \citenamefont {Bhatta~Pathak}, \citenamefont
  {Bhetuwal}, \citenamefont {Blaikie}, \citenamefont {Campagna}, \citenamefont
  {Camsonne}, \citenamefont {Cates}, \citenamefont {Chen}, \citenamefont
  {Clarke}, \citenamefont {Cornejo}, \citenamefont {Covrig~Dusa}, \citenamefont
  {Datta}, \citenamefont {Deshpande}, \citenamefont {Dutta}, \citenamefont
  {Feldman}, \citenamefont {Fuchey}, \citenamefont {Gal}, \citenamefont
  {Gaskell}, \citenamefont {Gautam}, \citenamefont {Gericke}, \citenamefont
  {Ghosh}, \citenamefont {Halilovic}, \citenamefont {Hansen}, \citenamefont
  {Hauenstein}, \citenamefont {Henry}, \citenamefont {Horowitz}, \citenamefont
  {Jantzi}, \citenamefont {Jian}, \citenamefont {Johnston}, \citenamefont
  {Jones}, \citenamefont {Karki}, \citenamefont {Katugampola}, \citenamefont
  {Keppel}, \citenamefont {King}, \citenamefont {King}, \citenamefont {Knauss},
  \citenamefont {Kumar}, \citenamefont {Kutz}, \citenamefont
  {Lashley-Colthirst}, \citenamefont {Leverick}, \citenamefont {Liu},
  \citenamefont {Liyange}, \citenamefont {Malace}, \citenamefont {Mammei},
  \citenamefont {Mammei}, \citenamefont {McCaughan}, \citenamefont {McNulty},
  \citenamefont {Meekins}, \citenamefont {Metts}, \citenamefont {Michaels},
  \citenamefont {Mondal}, \citenamefont {Napolitano}, \citenamefont {Narayan},
  \citenamefont {Nikolaev}, \citenamefont {Rashad}, \citenamefont {Owen},
  \citenamefont {Palatchi}, \citenamefont {Pan}, \citenamefont {Pandey},
  \citenamefont {Park}, \citenamefont {Paschke}, \citenamefont {Petrusky},
  \citenamefont {Pitt}, \citenamefont {Premathilake}, \citenamefont {Puckett},
  \citenamefont {Quinn}, \citenamefont {Radloff}, \citenamefont {Rahman},
  \citenamefont {Rathnayake}, \citenamefont {Reed}, \citenamefont {Reimer},
  \citenamefont {Richards}, \citenamefont {Riordan}, \citenamefont {Roblin},
  \citenamefont {Seeds}, \citenamefont {Shahinyan}, \citenamefont {Souder},
  \citenamefont {Tang}, \citenamefont {Thiel}, \citenamefont {Tian},
  \citenamefont {Urciuoli}, \citenamefont {Wertz}, \citenamefont
  {Wojtsekhowski}, \citenamefont {Yale}, \citenamefont {Ye}, \citenamefont
  {Yoon}, \citenamefont {Zec}, \citenamefont {Zhang}, \citenamefont {Zhang},\
  and\ \citenamefont {Zheng}}]{PREX2021}%
  \BibitemOpen
  \bibfield  {author} {\bibinfo {author} {\bibfnamefont {D.}~\bibnamefont
  {Adhikari}}, \bibinfo {author} {\bibfnamefont {H.}~\bibnamefont
  {Albataineh}}, \bibinfo {author} {\bibfnamefont {D.}~\bibnamefont {Androic}},
  \bibinfo {author} {\bibfnamefont {K.}~\bibnamefont {Aniol}}, \bibinfo
  {author} {\bibfnamefont {D.}~\bibnamefont {Armstrong}}, \bibinfo {author}
  {\bibfnamefont {T.}~\bibnamefont {Averett}}, \bibinfo {author} {\bibfnamefont
  {C.}~\bibnamefont {Ayerbe~Gayoso}}, \bibinfo {author} {\bibfnamefont
  {S.}~\bibnamefont {Barcus}}, \bibinfo {author} {\bibfnamefont
  {V.}~\bibnamefont {Bellini}}, \bibinfo {author} {\bibfnamefont
  {R.}~\bibnamefont {Beminiwattha}}, \bibinfo {author} {\bibfnamefont
  {J.}~\bibnamefont {Benesch}}, \bibinfo {author} {\bibfnamefont
  {H.}~\bibnamefont {Bhatt}}, \bibinfo {author} {\bibfnamefont
  {D.}~\bibnamefont {Bhatta~Pathak}}, \bibinfo {author} {\bibfnamefont
  {D.}~\bibnamefont {Bhetuwal}}, \bibinfo {author} {\bibfnamefont
  {B.}~\bibnamefont {Blaikie}}, \bibinfo {author} {\bibfnamefont
  {Q.}~\bibnamefont {Campagna}}, \bibinfo {author} {\bibfnamefont
  {A.}~\bibnamefont {Camsonne}}, \bibinfo {author} {\bibfnamefont
  {G.}~\bibnamefont {Cates}}, \bibinfo {author} {\bibfnamefont
  {Y.}~\bibnamefont {Chen}}, \bibinfo {author} {\bibfnamefont {C.}~\bibnamefont
  {Clarke}}, \bibinfo {author} {\bibfnamefont {J.}~\bibnamefont {Cornejo}},
  \bibinfo {author} {\bibfnamefont {S.}~\bibnamefont {Covrig~Dusa}}, \bibinfo
  {author} {\bibfnamefont {P.}~\bibnamefont {Datta}}, \bibinfo {author}
  {\bibfnamefont {A.}~\bibnamefont {Deshpande}}, \bibinfo {author}
  {\bibfnamefont {D.}~\bibnamefont {Dutta}}, \bibinfo {author} {\bibfnamefont
  {C.}~\bibnamefont {Feldman}}, \bibinfo {author} {\bibfnamefont
  {E.}~\bibnamefont {Fuchey}}, \bibinfo {author} {\bibfnamefont
  {C.}~\bibnamefont {Gal}}, \bibinfo {author} {\bibfnamefont {D.}~\bibnamefont
  {Gaskell}}, \bibinfo {author} {\bibfnamefont {T.}~\bibnamefont {Gautam}},
  \bibinfo {author} {\bibfnamefont {M.}~\bibnamefont {Gericke}}, \bibinfo
  {author} {\bibfnamefont {C.}~\bibnamefont {Ghosh}}, \bibinfo {author}
  {\bibfnamefont {I.}~\bibnamefont {Halilovic}}, \bibinfo {author}
  {\bibfnamefont {J.-O.}\ \bibnamefont {Hansen}}, \bibinfo {author}
  {\bibfnamefont {F.}~\bibnamefont {Hauenstein}}, \bibinfo {author}
  {\bibfnamefont {W.}~\bibnamefont {Henry}}, \bibinfo {author} {\bibfnamefont
  {C.}~\bibnamefont {Horowitz}}, \bibinfo {author} {\bibfnamefont
  {C.}~\bibnamefont {Jantzi}}, \bibinfo {author} {\bibfnamefont
  {S.}~\bibnamefont {Jian}}, \bibinfo {author} {\bibfnamefont {S.}~\bibnamefont
  {Johnston}}, \bibinfo {author} {\bibfnamefont {D.}~\bibnamefont {Jones}},
  \bibinfo {author} {\bibfnamefont {B.}~\bibnamefont {Karki}}, \bibinfo
  {author} {\bibfnamefont {S.}~\bibnamefont {Katugampola}}, \bibinfo {author}
  {\bibfnamefont {C.}~\bibnamefont {Keppel}}, \bibinfo {author} {\bibfnamefont
  {P.}~\bibnamefont {King}}, \bibinfo {author} {\bibfnamefont {D.}~\bibnamefont
  {King}}, \bibinfo {author} {\bibfnamefont {M.}~\bibnamefont {Knauss}},
  \bibinfo {author} {\bibfnamefont {K.}~\bibnamefont {Kumar}}, \bibinfo
  {author} {\bibfnamefont {T.}~\bibnamefont {Kutz}}, \bibinfo {author}
  {\bibfnamefont {N.}~\bibnamefont {Lashley-Colthirst}}, \bibinfo {author}
  {\bibfnamefont {G.}~\bibnamefont {Leverick}}, \bibinfo {author}
  {\bibfnamefont {H.}~\bibnamefont {Liu}}, \bibinfo {author} {\bibfnamefont
  {N.}~\bibnamefont {Liyange}}, \bibinfo {author} {\bibfnamefont
  {S.}~\bibnamefont {Malace}}, \bibinfo {author} {\bibfnamefont
  {R.}~\bibnamefont {Mammei}}, \bibinfo {author} {\bibfnamefont
  {J.}~\bibnamefont {Mammei}}, \bibinfo {author} {\bibfnamefont
  {M.}~\bibnamefont {McCaughan}}, \bibinfo {author} {\bibfnamefont
  {D.}~\bibnamefont {McNulty}}, \bibinfo {author} {\bibfnamefont
  {D.}~\bibnamefont {Meekins}}, \bibinfo {author} {\bibfnamefont
  {C.}~\bibnamefont {Metts}}, \bibinfo {author} {\bibfnamefont
  {R.}~\bibnamefont {Michaels}}, \bibinfo {author} {\bibfnamefont
  {M.}~\bibnamefont {Mondal}}, \bibinfo {author} {\bibfnamefont
  {J.}~\bibnamefont {Napolitano}}, \bibinfo {author} {\bibfnamefont
  {A.}~\bibnamefont {Narayan}}, \bibinfo {author} {\bibfnamefont
  {D.}~\bibnamefont {Nikolaev}}, \bibinfo {author} {\bibfnamefont
  {M.}~\bibnamefont {Rashad}}, \bibinfo {author} {\bibfnamefont
  {V.}~\bibnamefont {Owen}}, \bibinfo {author} {\bibfnamefont {C.}~\bibnamefont
  {Palatchi}}, \bibinfo {author} {\bibfnamefont {J.}~\bibnamefont {Pan}},
  \bibinfo {author} {\bibfnamefont {B.}~\bibnamefont {Pandey}}, \bibinfo
  {author} {\bibfnamefont {S.}~\bibnamefont {Park}}, \bibinfo {author}
  {\bibfnamefont {K.}~\bibnamefont {Paschke}}, \bibinfo {author} {\bibfnamefont
  {M.}~\bibnamefont {Petrusky}}, \bibinfo {author} {\bibfnamefont
  {M.}~\bibnamefont {Pitt}}, \bibinfo {author} {\bibfnamefont {S.}~\bibnamefont
  {Premathilake}}, \bibinfo {author} {\bibfnamefont {A.}~\bibnamefont
  {Puckett}}, \bibinfo {author} {\bibfnamefont {B.}~\bibnamefont {Quinn}},
  \bibinfo {author} {\bibfnamefont {R.}~\bibnamefont {Radloff}}, \bibinfo
  {author} {\bibfnamefont {S.}~\bibnamefont {Rahman}}, \bibinfo {author}
  {\bibfnamefont {A.}~\bibnamefont {Rathnayake}}, \bibinfo {author}
  {\bibfnamefont {B.}~\bibnamefont {Reed}}, \bibinfo {author} {\bibfnamefont
  {P.}~\bibnamefont {Reimer}}, \bibinfo {author} {\bibfnamefont
  {R.}~\bibnamefont {Richards}}, \bibinfo {author} {\bibfnamefont
  {S.}~\bibnamefont {Riordan}}, \bibinfo {author} {\bibfnamefont
  {Y.}~\bibnamefont {Roblin}}, \bibinfo {author} {\bibfnamefont
  {S.}~\bibnamefont {Seeds}}, \bibinfo {author} {\bibfnamefont
  {A.}~\bibnamefont {Shahinyan}}, \bibinfo {author} {\bibfnamefont
  {P.}~\bibnamefont {Souder}}, \bibinfo {author} {\bibfnamefont
  {L.}~\bibnamefont {Tang}}, \bibinfo {author} {\bibfnamefont {M.}~\bibnamefont
  {Thiel}}, \bibinfo {author} {\bibfnamefont {Y.}~\bibnamefont {Tian}},
  \bibinfo {author} {\bibfnamefont {G.}~\bibnamefont {Urciuoli}}, \bibinfo
  {author} {\bibfnamefont {E.}~\bibnamefont {Wertz}}, \bibinfo {author}
  {\bibfnamefont {B.}~\bibnamefont {Wojtsekhowski}}, \bibinfo {author}
  {\bibfnamefont {B.}~\bibnamefont {Yale}}, \bibinfo {author} {\bibfnamefont
  {T.}~\bibnamefont {Ye}}, \bibinfo {author} {\bibfnamefont {A.}~\bibnamefont
  {Yoon}}, \bibinfo {author} {\bibfnamefont {A.}~\bibnamefont {Zec}}, \bibinfo
  {author} {\bibfnamefont {W.}~\bibnamefont {Zhang}}, \bibinfo {author}
  {\bibfnamefont {J.}~\bibnamefont {Zhang}},\ and\ \bibinfo {author}
  {\bibfnamefont {X.}~\bibnamefont {Zheng}},\ }\bibfield  {title} {\bibinfo
  {title} {{Accurate Determination of the Neutron Skin Thickness of 208Pb
  through Parity-Violation in Electron Scattering}},\ }\href
  {https://doi.org/10.1103/PhysRevLett.126.172502} {\bibfield  {journal}
  {\bibinfo  {journal} {Physical Review Letters}\ }\textbf {\bibinfo {volume}
  {126}},\ \bibinfo {pages} {172502} (\bibinfo {year} {2021})}\BibitemShut
  {NoStop}%
\bibitem [{\citenamefont {Lattimer}\ and\ \citenamefont
  {Prakash}(2000)}]{Lattimer2000}%
  \BibitemOpen
  \bibfield  {author} {\bibinfo {author} {\bibfnamefont {J.~M.}\ \bibnamefont
  {Lattimer}}\ and\ \bibinfo {author} {\bibfnamefont {M.}~\bibnamefont
  {Prakash}},\ }\bibfield  {title} {\bibinfo {title} {{Nuclear matter and its
  role in supernovae, neutron stars and compact object binary mergers}},\
  }\href {https://doi.org/https://doi.org/10.1016/S0370-1573(00)00019-3}
  {\bibfield  {journal} {\bibinfo  {journal} {Physics Reports}\ }\textbf
  {\bibinfo {volume} {333-334}},\ \bibinfo {pages} {121} (\bibinfo {year}
  {2000})}\BibitemShut {NoStop}%
\bibitem [{\citenamefont {Reed}\ \emph {et~al.}(2021)\citenamefont {Reed},
  \citenamefont {Fattoyev}, \citenamefont {Horowitz},\ and\ \citenamefont
  {Piekarewicz}}]{Reed2021}%
  \BibitemOpen
  \bibfield  {author} {\bibinfo {author} {\bibfnamefont {B.~T.}\ \bibnamefont
  {Reed}}, \bibinfo {author} {\bibfnamefont {F.}~\bibnamefont {Fattoyev}},
  \bibinfo {author} {\bibfnamefont {C.}~\bibnamefont {Horowitz}},\ and\
  \bibinfo {author} {\bibfnamefont {J.}~\bibnamefont {Piekarewicz}},\
  }\bibfield  {title} {\bibinfo {title} {{Implications of PREX-2 on the
  Equation of State of Neutron-Rich Matter}},\ }\href
  {https://doi.org/10.1103/PhysRevLett.126.172503} {\bibfield  {journal}
  {\bibinfo  {journal} {Physical Review Letters}\ }\textbf {\bibinfo {volume}
  {126}},\ \bibinfo {pages} {172503} (\bibinfo {year} {2021})}\BibitemShut
  {NoStop}%
\bibitem [{\citenamefont {Egelhof}\ \emph {et~al.}(2002)\citenamefont
  {Egelhof}, \citenamefont {Alkhazov}, \citenamefont {Andronenko},
  \citenamefont {Bauchet}, \citenamefont {Dobrovolsky}, \citenamefont {Fritz},
  \citenamefont {Gavrilov}, \citenamefont {Geissel}, \citenamefont {Gross},
  \citenamefont {Khanzadeev}, \citenamefont {Korolev}, \citenamefont {Kraus},
  \citenamefont {Lobodenko}, \citenamefont {M{\"{u}}nzenberg}, \citenamefont
  {Mutterer}, \citenamefont {Neumaier}, \citenamefont {Sch{\"{a}}fer},
  \citenamefont {Scheidenberger}, \citenamefont {Seliverstov}, \citenamefont
  {Timofeev}, \citenamefont {Vorobyov},\ and\ \citenamefont
  {Yatsoura}}]{Egelhof2002}%
  \BibitemOpen
  \bibfield  {author} {\bibinfo {author} {\bibfnamefont {P.}~\bibnamefont
  {Egelhof}}, \bibinfo {author} {\bibfnamefont {G.~D.}\ \bibnamefont
  {Alkhazov}}, \bibinfo {author} {\bibfnamefont {M.~N.}\ \bibnamefont
  {Andronenko}}, \bibinfo {author} {\bibfnamefont {A.}~\bibnamefont {Bauchet}},
  \bibinfo {author} {\bibfnamefont {A.~V.}\ \bibnamefont {Dobrovolsky}},
  \bibinfo {author} {\bibfnamefont {S.}~\bibnamefont {Fritz}}, \bibinfo
  {author} {\bibfnamefont {G.~E.}\ \bibnamefont {Gavrilov}}, \bibinfo {author}
  {\bibfnamefont {H.}~\bibnamefont {Geissel}}, \bibinfo {author} {\bibfnamefont
  {C.}~\bibnamefont {Gross}}, \bibinfo {author} {\bibfnamefont {A.~V.}\
  \bibnamefont {Khanzadeev}}, \bibinfo {author} {\bibfnamefont {G.~A.}\
  \bibnamefont {Korolev}}, \bibinfo {author} {\bibfnamefont {G.}~\bibnamefont
  {Kraus}}, \bibinfo {author} {\bibfnamefont {A.~A.}\ \bibnamefont
  {Lobodenko}}, \bibinfo {author} {\bibfnamefont {G.}~\bibnamefont
  {M{\"{u}}nzenberg}}, \bibinfo {author} {\bibfnamefont {M.}~\bibnamefont
  {Mutterer}}, \bibinfo {author} {\bibfnamefont {S.~R.}\ \bibnamefont
  {Neumaier}}, \bibinfo {author} {\bibfnamefont {T.}~\bibnamefont
  {Sch{\"{a}}fer}}, \bibinfo {author} {\bibfnamefont {C.}~\bibnamefont
  {Scheidenberger}}, \bibinfo {author} {\bibfnamefont {D.~M.}\ \bibnamefont
  {Seliverstov}}, \bibinfo {author} {\bibfnamefont {N.~A.}\ \bibnamefont
  {Timofeev}}, \bibinfo {author} {\bibfnamefont {A.~A.}\ \bibnamefont
  {Vorobyov}},\ and\ \bibinfo {author} {\bibfnamefont {V.~I.}\ \bibnamefont
  {Yatsoura}},\ }\bibfield  {title} {\bibinfo {title} {{Nuclear-matter
  distributions of halo nuclei from elastic proton scattering in inverse
  kinematics}},\ }\href {https://doi.org/10.1140/epja/i2001-10219-7} {\bibfield
   {journal} {\bibinfo  {journal} {The European Physical Journal A}\ }\textbf
  {\bibinfo {volume} {15}},\ \bibinfo {pages} {27} (\bibinfo {year}
  {2002})}\BibitemShut {NoStop}%
\bibitem [{\citenamefont {Dobrovolsky}\ \emph {et~al.}(2006)\citenamefont
  {Dobrovolsky}, \citenamefont {Alkhazov}, \citenamefont {Andronenko},
  \citenamefont {Bauchet}, \citenamefont {Egelhof}, \citenamefont {Fritz},
  \citenamefont {Geissel}, \citenamefont {Gross}, \citenamefont {Khanzadeev},
  \citenamefont {Korolev}, \citenamefont {Kraus}, \citenamefont {Lobodenko},
  \citenamefont {M{\"{u}}nzenberg}, \citenamefont {Mutterer}, \citenamefont
  {Neumaier}, \citenamefont {Sch{\"{a}}fer}, \citenamefont {Scheidenberger},
  \citenamefont {Seliverstov}, \citenamefont {Timofeev}, \citenamefont
  {Vorobyov},\ and\ \citenamefont {Yatsoura}}]{Dobrovolsky2006}%
  \BibitemOpen
  \bibfield  {author} {\bibinfo {author} {\bibfnamefont {A.~V.}\ \bibnamefont
  {Dobrovolsky}}, \bibinfo {author} {\bibfnamefont {G.~D.}\ \bibnamefont
  {Alkhazov}}, \bibinfo {author} {\bibfnamefont {M.~N.}\ \bibnamefont
  {Andronenko}}, \bibinfo {author} {\bibfnamefont {A.}~\bibnamefont {Bauchet}},
  \bibinfo {author} {\bibfnamefont {P.}~\bibnamefont {Egelhof}}, \bibinfo
  {author} {\bibfnamefont {S.}~\bibnamefont {Fritz}}, \bibinfo {author}
  {\bibfnamefont {H.}~\bibnamefont {Geissel}}, \bibinfo {author} {\bibfnamefont
  {C.}~\bibnamefont {Gross}}, \bibinfo {author} {\bibfnamefont {A.~V.}\
  \bibnamefont {Khanzadeev}}, \bibinfo {author} {\bibfnamefont {G.~A.}\
  \bibnamefont {Korolev}}, \bibinfo {author} {\bibfnamefont {G.}~\bibnamefont
  {Kraus}}, \bibinfo {author} {\bibfnamefont {A.~A.}\ \bibnamefont
  {Lobodenko}}, \bibinfo {author} {\bibfnamefont {G.}~\bibnamefont
  {M{\"{u}}nzenberg}}, \bibinfo {author} {\bibfnamefont {M.}~\bibnamefont
  {Mutterer}}, \bibinfo {author} {\bibfnamefont {S.~R.}\ \bibnamefont
  {Neumaier}}, \bibinfo {author} {\bibfnamefont {T.}~\bibnamefont
  {Sch{\"{a}}fer}}, \bibinfo {author} {\bibfnamefont {C.}~\bibnamefont
  {Scheidenberger}}, \bibinfo {author} {\bibfnamefont {D.~M.}\ \bibnamefont
  {Seliverstov}}, \bibinfo {author} {\bibfnamefont {N.~A.}\ \bibnamefont
  {Timofeev}}, \bibinfo {author} {\bibfnamefont {A.~A.}\ \bibnamefont
  {Vorobyov}},\ and\ \bibinfo {author} {\bibfnamefont {V.~I.}\ \bibnamefont
  {Yatsoura}},\ }\bibfield  {title} {\bibinfo {title} {{Study of the nuclear
  matter distribution in neutron-rich Li isotopes}},\ }\href
  {https://doi.org/https://doi.org/10.1016/j.nuclphysa.2005.11.016} {\bibfield
  {journal} {\bibinfo  {journal} {Nuclear Physics A}\ }\textbf {\bibinfo
  {volume} {766}},\ \bibinfo {pages} {1} (\bibinfo {year} {2006})}\BibitemShut
  {NoStop}%
\bibitem [{\citenamefont {Erler}\ \emph {et~al.}(2012)\citenamefont {Erler},
  \citenamefont {Birge}, \citenamefont {Kortelainen}, \citenamefont
  {Nazarewicz}, \citenamefont {Olsen}, \citenamefont {Perhac},\ and\
  \citenamefont {Stoitsov}}]{Erler2012}%
  \BibitemOpen
  \bibfield  {author} {\bibinfo {author} {\bibfnamefont {J.}~\bibnamefont
  {Erler}}, \bibinfo {author} {\bibfnamefont {N.}~\bibnamefont {Birge}},
  \bibinfo {author} {\bibfnamefont {M.}~\bibnamefont {Kortelainen}}, \bibinfo
  {author} {\bibfnamefont {W.}~\bibnamefont {Nazarewicz}}, \bibinfo {author}
  {\bibfnamefont {E.}~\bibnamefont {Olsen}}, \bibinfo {author} {\bibfnamefont
  {A.~M.}\ \bibnamefont {Perhac}},\ and\ \bibinfo {author} {\bibfnamefont
  {M.}~\bibnamefont {Stoitsov}},\ }\bibfield  {title} {\bibinfo {title} {{The
  limits of the nuclear landscape}},\ }\href
  {https://doi.org/10.1038/nature11188} {\bibfield  {journal} {\bibinfo
  {journal} {Nature}\ }\textbf {\bibinfo {volume} {486}},\ \bibinfo {pages}
  {509} (\bibinfo {year} {2012})}\BibitemShut {NoStop}%
\bibitem [{\citenamefont {Tsunoda}\ \emph {et~al.}(2020)\citenamefont
  {Tsunoda}, \citenamefont {Otsuka}, \citenamefont {Takayanagi}, \citenamefont
  {Shimizu}, \citenamefont {Suzuki}, \citenamefont {Utsuno}, \citenamefont
  {Yoshida},\ and\ \citenamefont {Ueno}}]{Tsunoda2020}%
  \BibitemOpen
  \bibfield  {author} {\bibinfo {author} {\bibfnamefont {N.}~\bibnamefont
  {Tsunoda}}, \bibinfo {author} {\bibfnamefont {T.}~\bibnamefont {Otsuka}},
  \bibinfo {author} {\bibfnamefont {K.}~\bibnamefont {Takayanagi}}, \bibinfo
  {author} {\bibfnamefont {N.}~\bibnamefont {Shimizu}}, \bibinfo {author}
  {\bibfnamefont {T.}~\bibnamefont {Suzuki}}, \bibinfo {author} {\bibfnamefont
  {Y.}~\bibnamefont {Utsuno}}, \bibinfo {author} {\bibfnamefont
  {S.}~\bibnamefont {Yoshida}},\ and\ \bibinfo {author} {\bibfnamefont
  {H.}~\bibnamefont {Ueno}},\ }\bibfield  {title} {\bibinfo {title} {{The
  impact of nuclear shape on the emergence of the neutron dripline}},\ }\href
  {https://doi.org/10.1038/s41586-020-2848-x} {\bibfield  {journal} {\bibinfo
  {journal} {Nature}\ }\textbf {\bibinfo {volume} {587}},\ \bibinfo {pages}
  {66} (\bibinfo {year} {2020})}\BibitemShut {NoStop}%
\bibitem [{\citenamefont {Wang}\ \emph {et~al.}(2021)\citenamefont {Wang},
  \citenamefont {Huang}, \citenamefont {Kondev}, \citenamefont {Audi},\ and\
  \citenamefont {Naimi}}]{AME2020}%
  \BibitemOpen
  \bibfield  {author} {\bibinfo {author} {\bibfnamefont {M.}~\bibnamefont
  {Wang}}, \bibinfo {author} {\bibfnamefont {W.~J.}\ \bibnamefont {Huang}},
  \bibinfo {author} {\bibfnamefont {F.~G.}\ \bibnamefont {Kondev}}, \bibinfo
  {author} {\bibfnamefont {G.}~\bibnamefont {Audi}},\ and\ \bibinfo {author}
  {\bibfnamefont {S.}~\bibnamefont {Naimi}},\ }\bibfield  {title} {\bibinfo
  {title} {{The AME 2020 atomic mass evaluation (II). Tables, graphs and
  references*}},\ }\href {https://doi.org/10.1088/1674-1137/abddaf} {\bibfield
  {journal} {\bibinfo  {journal} {Chinese Physics C}\ }\textbf {\bibinfo
  {volume} {45}},\ \bibinfo {pages} {30003} (\bibinfo {year}
  {2021})}\BibitemShut {NoStop}%
\bibitem [{\citenamefont {Wang}\ \emph {et~al.}(2012)\citenamefont {Wang},
  \citenamefont {Liu}, \citenamefont {Zhang}, \citenamefont {Cao},\ and\
  \citenamefont {Jia}}]{Wang2012Vibr}%
  \BibitemOpen
  \bibfield  {author} {\bibinfo {author} {\bibfnamefont {P.~Q.}\ \bibnamefont
  {Wang}}, \bibinfo {author} {\bibfnamefont {J.~Y.}\ \bibnamefont {Liu}},
  \bibinfo {author} {\bibfnamefont {L.~H.}\ \bibnamefont {Zhang}}, \bibinfo
  {author} {\bibfnamefont {S.~Y.}\ \bibnamefont {Cao}},\ and\ \bibinfo {author}
  {\bibfnamefont {C.~S.}\ \bibnamefont {Jia}},\ }\bibfield  {title} {\bibinfo
  {title} {{Improved expressions for the Schi{\"{o}}berg potential energy
  models for diatomic molecules}},\ }\href
  {https://doi.org/10.1016/j.jms.2012.07.001} {\bibfield  {journal} {\bibinfo
  {journal} {Journal of Molecular Spectroscopy}\ }\textbf {\bibinfo {volume}
  {278}},\ \bibinfo {pages} {23} (\bibinfo {year} {2012})}\BibitemShut
  {NoStop}%
\bibitem [{\citenamefont {Hutzler}\ \emph {et~al.}(2011)\citenamefont
  {Hutzler}, \citenamefont {Parsons}, \citenamefont {Gurevich}, \citenamefont
  {Hess}, \citenamefont {Petrik}, \citenamefont {Spaun}, \citenamefont {Vutha},
  \citenamefont {DeMille}, \citenamefont {Gabrielse},\ and\ \citenamefont
  {Doyle}}]{Hutzler2011}%
  \BibitemOpen
  \bibfield  {author} {\bibinfo {author} {\bibfnamefont {N.~R.}\ \bibnamefont
  {Hutzler}}, \bibinfo {author} {\bibfnamefont {M.~F.}\ \bibnamefont
  {Parsons}}, \bibinfo {author} {\bibfnamefont {Y.~V.}\ \bibnamefont
  {Gurevich}}, \bibinfo {author} {\bibfnamefont {P.~W.}\ \bibnamefont {Hess}},
  \bibinfo {author} {\bibfnamefont {E.}~\bibnamefont {Petrik}}, \bibinfo
  {author} {\bibfnamefont {B.}~\bibnamefont {Spaun}}, \bibinfo {author}
  {\bibfnamefont {A.~C.}\ \bibnamefont {Vutha}}, \bibinfo {author}
  {\bibfnamefont {D.}~\bibnamefont {DeMille}}, \bibinfo {author} {\bibfnamefont
  {G.}~\bibnamefont {Gabrielse}},\ and\ \bibinfo {author} {\bibfnamefont
  {J.~M.}\ \bibnamefont {Doyle}},\ }\bibfield  {title} {\bibinfo {title} {{A
  cryogenic beam of refractory, chemically reactive molecules with expansion
  cooling}},\ }\href {https://doi.org/10.1039/C1CP20901A} {\bibfield  {journal}
  {\bibinfo  {journal} {Physical Chemistry Chemical Physics}\ }\textbf
  {\bibinfo {volume} {13}},\ \bibinfo {pages} {18976} (\bibinfo {year}
  {2011})}\BibitemShut {NoStop}%
\bibitem [{\citenamefont {{A Del Sol Mesa}}\ \emph {et~al.}(1998)\citenamefont
  {{A Del Sol Mesa}}, \citenamefont {{C Quesne}},\ and\ \citenamefont {{Yu F
  Smirnov}}}]{DelSolMesa1998}%
  \BibitemOpen
  \bibfield  {author} {\bibinfo {author} {\bibnamefont {{A Del Sol Mesa}}},
  \bibinfo {author} {\bibnamefont {{C Quesne}}},\ and\ \bibinfo {author}
  {\bibnamefont {{Yu F Smirnov}}},\ }\bibfield  {title} {\bibinfo {title}
  {{Generalized Morse potential: Symmetry and satellite potentials}},\ }\href
  {https://doi.org/10.1088/0305-4470/31/1/028} {\bibfield  {journal} {\bibinfo
  {journal} {Journal of Physics A: Mathematical and General}\ }\textbf
  {\bibinfo {volume} {31}},\ \bibinfo {pages} {321} (\bibinfo {year}
  {1998})}\BibitemShut {NoStop}%
\bibitem [{\citenamefont {Abu-Shady}\ and\ \citenamefont
  {Khokha}(2023)}]{AbuShady2023}%
  \BibitemOpen
  \bibfield  {author} {\bibinfo {author} {\bibfnamefont {M.}~\bibnamefont
  {Abu-Shady}}\ and\ \bibinfo {author} {\bibfnamefont {E.~M.}\ \bibnamefont
  {Khokha}},\ }\bibfield  {title} {\bibinfo {title} {{A precise estimation for
  vibrational energies of diatomic molecules using the improved Rosen–Morse
  potential}},\ }\href {https://doi.org/10.1038/s41598-023-37888-2} {\bibfield
  {journal} {\bibinfo  {journal} {Scientific Reports}\ }\textbf {\bibinfo
  {volume} {13}},\ \bibinfo {pages} {11578} (\bibinfo {year}
  {2023})}\BibitemShut {NoStop}%
\bibitem [{\citenamefont {Dunham}(1932)}]{Dunham1932}%
  \BibitemOpen
  \bibfield  {author} {\bibinfo {author} {\bibfnamefont {J.~L.}\ \bibnamefont
  {Dunham}},\ }\bibfield  {title} {\bibinfo {title} {{The Energy Levels of a
  Rotating Vibrator}},\ }\href {https://doi.org/10.1103/PhysRev.41.721}
  {\bibfield  {journal} {\bibinfo  {journal} {Phys. Rev.}\ }\textbf {\bibinfo
  {volume} {41}},\ \bibinfo {pages} {721} (\bibinfo {year} {1932})}\BibitemShut
  {NoStop}%
\bibitem [{\citenamefont {Schlembach}\ and\ \citenamefont
  {Tiemann}(1982)}]{Schlembach1982}%
  \BibitemOpen
  \bibfield  {author} {\bibinfo {author} {\bibfnamefont {J.}~\bibnamefont
  {Schlembach}}\ and\ \bibinfo {author} {\bibfnamefont {E.}~\bibnamefont
  {Tiemann}},\ }\bibfield  {title} {\bibinfo {title} {{Isotopic field shift of
  the rotational energy of the Pb-chalcogenides and Tl-Halides}},\ }\href
  {https://doi.org/https://doi.org/10.1016/0301-0104(82)85077-5} {\bibfield
  {journal} {\bibinfo  {journal} {Chemical Physics}\ }\textbf {\bibinfo
  {volume} {68}},\ \bibinfo {pages} {21} (\bibinfo {year} {1982})}\BibitemShut
  {NoStop}%
\bibitem [{\citenamefont {Knecht}\ and\ \citenamefont
  {Saue}(2012)}]{Knecht2012}%
  \BibitemOpen
  \bibfield  {author} {\bibinfo {author} {\bibfnamefont {S.}~\bibnamefont
  {Knecht}}\ and\ \bibinfo {author} {\bibfnamefont {T.}~\bibnamefont {Saue}},\
  }\bibfield  {title} {\bibinfo {title} {{Nuclear size effects in rotational
  spectra: A tale with a twist}},\ }\href
  {https://doi.org/https://doi.org/10.1016/j.chemphys.2011.10.030} {\bibfield
  {journal} {\bibinfo  {journal} {Chemical Physics}\ }\textbf {\bibinfo
  {volume} {401}},\ \bibinfo {pages} {103} (\bibinfo {year}
  {2012})}\BibitemShut {NoStop}%
\bibitem [{\citenamefont {Athanasakis-Kaklamanakis}\ \emph
  {et~al.}(2023)\citenamefont {Athanasakis-Kaklamanakis}, \citenamefont
  {Wilkins}, \citenamefont {Breier},\ and\ \citenamefont
  {Neyens}}]{AthanasakisKaklamanakis2023kingplot}%
  \BibitemOpen
  \bibfield  {author} {\bibinfo {author} {\bibfnamefont {M.}~\bibnamefont
  {Athanasakis-Kaklamanakis}}, \bibinfo {author} {\bibfnamefont {S.~G.}\
  \bibnamefont {Wilkins}}, \bibinfo {author} {\bibfnamefont {A.~A.}\
  \bibnamefont {Breier}},\ and\ \bibinfo {author} {\bibfnamefont
  {G.}~\bibnamefont {Neyens}},\ }\bibfield  {title} {\bibinfo {title}
  {{King-Plot Analysis of Isotope Shifts in Simple Diatomic Molecules}},\
  }\href {https://doi.org/10.1103/PhysRevX.13.011015} {\bibfield  {journal}
  {\bibinfo  {journal} {Phys. Rev. X}\ }\textbf {\bibinfo {volume} {13}},\
  \bibinfo {pages} {11015} (\bibinfo {year} {2023})}\BibitemShut {NoStop}%
\bibitem [{\citenamefont {Skripnikov}\ \emph {et~al.}(2021)\citenamefont
  {Skripnikov}, \citenamefont {Chubukov},\ and\ \citenamefont
  {Shakhova}}]{Skripnikov2021b}%
  \BibitemOpen
  \bibfield  {author} {\bibinfo {author} {\bibfnamefont {L.~V.}\ \bibnamefont
  {Skripnikov}}, \bibinfo {author} {\bibfnamefont {D.~V.}\ \bibnamefont
  {Chubukov}},\ and\ \bibinfo {author} {\bibfnamefont {V.~M.}\ \bibnamefont
  {Shakhova}},\ }\bibfield  {title} {\bibinfo {title} {{The role of QED effects
  in transition energies of heavy-atom alkaline earth monofluoride molecules: A
  theoretical study of Ba+, BaF, RaF, and E120F}},\ }\href
  {https://doi.org/10.1063/5.0068267} {\bibfield  {journal} {\bibinfo
  {journal} {The Journal of Chemical Physics}\ }\textbf {\bibinfo {volume}
  {155}},\ \bibinfo {pages} {144103} (\bibinfo {year} {2021})}\BibitemShut
  {NoStop}%
\bibitem [{\citenamefont {Navr{\'{a}}til}\ \emph {et~al.}(2016)\citenamefont
  {Navr{\'{a}}til}, \citenamefont {Quaglioni}, \citenamefont {Hupin},
  \citenamefont {Romero-Redondo},\ and\ \citenamefont {Calci}}]{Navratil2016}%
  \BibitemOpen
  \bibfield  {author} {\bibinfo {author} {\bibfnamefont {P.}~\bibnamefont
  {Navr{\'{a}}til}}, \bibinfo {author} {\bibfnamefont {S.}~\bibnamefont
  {Quaglioni}}, \bibinfo {author} {\bibfnamefont {G.}~\bibnamefont {Hupin}},
  \bibinfo {author} {\bibfnamefont {C.}~\bibnamefont {Romero-Redondo}},\ and\
  \bibinfo {author} {\bibfnamefont {A.}~\bibnamefont {Calci}},\ }\bibfield
  {title} {\bibinfo {title} {{Unified ab initioapproaches to nuclear structure
  and reactions}},\ }\href {https://doi.org/10.1088/0031-8949/91/5/053002}
  {\bibfield  {journal} {\bibinfo  {journal} {Physica Scripta}\ }\textbf
  {\bibinfo {volume} {91}},\ \bibinfo {pages} {53002} (\bibinfo {year}
  {2016})}\BibitemShut {NoStop}%
\bibitem [{\citenamefont {Hergert}(2020)}]{Hergert2020}%
  \BibitemOpen
  \bibfield  {author} {\bibinfo {author} {\bibfnamefont {H.}~\bibnamefont
  {Hergert}},\ }\bibfield  {title} {\bibinfo {title} {{A Guided Tour of ab
  initio Nuclear Many-Body Theory}},\ }\bibfield  {journal} {\bibinfo
  {journal} {Frontiers in Physics}\ }\textbf {\bibinfo {volume} {8}},\ \href
  {https://doi.org/10.3389/fphy.2020.00379} {10.3389/fphy.2020.00379} (\bibinfo
  {year} {2020})\BibitemShut {NoStop}%
\bibitem [{\citenamefont {Collaboration}\ \emph {et~al.}(2014)\citenamefont
  {Collaboration}, \citenamefont {Beane}, \citenamefont {Chang}, \citenamefont
  {Cohen}, \citenamefont {Detmold}, \citenamefont {Lin}, \citenamefont
  {Orginos}, \citenamefont {Parre{\~{n}}o}, \citenamefont {Savage},\ and\
  \citenamefont {Tiburzi}}]{Beane2014NPLQCD}%
  \BibitemOpen
  \bibfield  {author} {\bibinfo {author} {\bibfnamefont {N.}~\bibnamefont
  {Collaboration}}, \bibinfo {author} {\bibfnamefont {S.}~\bibnamefont
  {Beane}}, \bibinfo {author} {\bibfnamefont {E.}~\bibnamefont {Chang}},
  \bibinfo {author} {\bibfnamefont {S.}~\bibnamefont {Cohen}}, \bibinfo
  {author} {\bibfnamefont {W.}~\bibnamefont {Detmold}}, \bibinfo {author}
  {\bibfnamefont {H.}~\bibnamefont {Lin}}, \bibinfo {author} {\bibfnamefont
  {K.}~\bibnamefont {Orginos}}, \bibinfo {author} {\bibfnamefont
  {A.}~\bibnamefont {Parre{\~{n}}o}}, \bibinfo {author} {\bibfnamefont
  {M.}~\bibnamefont {Savage}},\ and\ \bibinfo {author} {\bibfnamefont
  {B.}~\bibnamefont {Tiburzi}},\ }\bibfield  {title} {\bibinfo {title}
  {{Magnetic Moments of Light Nuclei from Lattice Quantum Chromodynamics}},\
  }\href {https://doi.org/10.1103/PhysRevLett.113.252001} {\bibfield  {journal}
  {\bibinfo  {journal} {Physical Review Letters}\ }\textbf {\bibinfo {volume}
  {113}},\ \bibinfo {pages} {252001} (\bibinfo {year} {2014})}\BibitemShut
  {NoStop}%
\bibitem [{\citenamefont {Collaboration}\ \emph {et~al.}(2015)\citenamefont
  {Collaboration}, \citenamefont {Chang}, \citenamefont {Detmold},
  \citenamefont {Orginos}, \citenamefont {Parre{\~{n}}o}, \citenamefont
  {Savage}, \citenamefont {Tiburzi},\ and\ \citenamefont
  {Beane}}]{Chang2015NPLQCD}%
  \BibitemOpen
  \bibfield  {author} {\bibinfo {author} {\bibfnamefont {N.}~\bibnamefont
  {Collaboration}}, \bibinfo {author} {\bibfnamefont {E.}~\bibnamefont
  {Chang}}, \bibinfo {author} {\bibfnamefont {W.}~\bibnamefont {Detmold}},
  \bibinfo {author} {\bibfnamefont {K.}~\bibnamefont {Orginos}}, \bibinfo
  {author} {\bibfnamefont {A.}~\bibnamefont {Parre{\~{n}}o}}, \bibinfo {author}
  {\bibfnamefont {M.~J.}\ \bibnamefont {Savage}}, \bibinfo {author}
  {\bibfnamefont {B.~C.}\ \bibnamefont {Tiburzi}},\ and\ \bibinfo {author}
  {\bibfnamefont {S.~R.}\ \bibnamefont {Beane}},\ }\bibfield  {title} {\bibinfo
  {title} {{Magnetic structure of light nuclei from lattice QCD}},\ }\href
  {https://doi.org/10.1103/PhysRevD.92.114502} {\bibfield  {journal} {\bibinfo
  {journal} {Physical Review D}\ }\textbf {\bibinfo {volume} {92}},\ \bibinfo
  {pages} {114502} (\bibinfo {year} {2015})}\BibitemShut {NoStop}%
\bibitem [{\citenamefont {Collaboration}\ \emph {et~al.}(2021)\citenamefont
  {Collaboration}, \citenamefont {Parre{\~{n}}o}, \citenamefont {Shanahan},
  \citenamefont {Wagman}, \citenamefont {Winter}, \citenamefont {Chang},
  \citenamefont {Detmold},\ and\ \citenamefont {Illa}}]{Parreno2021NPLQCD}%
  \BibitemOpen
  \bibfield  {author} {\bibinfo {author} {\bibfnamefont {N.}~\bibnamefont
  {Collaboration}}, \bibinfo {author} {\bibfnamefont {A.}~\bibnamefont
  {Parre{\~{n}}o}}, \bibinfo {author} {\bibfnamefont {P.~E.}\ \bibnamefont
  {Shanahan}}, \bibinfo {author} {\bibfnamefont {M.~L.}\ \bibnamefont
  {Wagman}}, \bibinfo {author} {\bibfnamefont {F.}~\bibnamefont {Winter}},
  \bibinfo {author} {\bibfnamefont {E.}~\bibnamefont {Chang}}, \bibinfo
  {author} {\bibfnamefont {W.}~\bibnamefont {Detmold}},\ and\ \bibinfo {author}
  {\bibfnamefont {M.}~\bibnamefont {Illa}},\ }\bibfield  {title} {\bibinfo
  {title} {{Axial charge of the triton from lattice QCD}},\ }\href
  {https://doi.org/10.1103/PhysRevD.103.074511} {\bibfield  {journal} {\bibinfo
   {journal} {Physical Review D}\ }\textbf {\bibinfo {volume} {103}},\ \bibinfo
  {pages} {74511} (\bibinfo {year} {2021})}\BibitemShut {NoStop}%
\bibitem [{\citenamefont {Gao}\ and\ \citenamefont
  {Vanderhaeghen}(2022)}]{Gao2022ProtonRadius}%
  \BibitemOpen
  \bibfield  {author} {\bibinfo {author} {\bibfnamefont {H.}~\bibnamefont
  {Gao}}\ and\ \bibinfo {author} {\bibfnamefont {M.}~\bibnamefont
  {Vanderhaeghen}},\ }\bibfield  {title} {\bibinfo {title} {{The proton charge
  radius}},\ }\href {https://doi.org/10.1103/RevModPhys.94.015002} {\bibfield
  {journal} {\bibinfo  {journal} {Reviews of Modern Physics}\ }\textbf
  {\bibinfo {volume} {94}},\ \bibinfo {pages} {15002} (\bibinfo {year}
  {2022})}\BibitemShut {NoStop}%
\bibitem [{\citenamefont {Koelemeij}\ \emph {et~al.}(2007)\citenamefont
  {Koelemeij}, \citenamefont {Roth}, \citenamefont {Wicht}, \citenamefont
  {Ernsting},\ and\ \citenamefont {Schiller}}]{Koelemeij2007}%
  \BibitemOpen
  \bibfield  {author} {\bibinfo {author} {\bibfnamefont {J.~C.}\ \bibnamefont
  {Koelemeij}}, \bibinfo {author} {\bibfnamefont {B.}~\bibnamefont {Roth}},
  \bibinfo {author} {\bibfnamefont {A.}~\bibnamefont {Wicht}}, \bibinfo
  {author} {\bibfnamefont {I.}~\bibnamefont {Ernsting}},\ and\ \bibinfo
  {author} {\bibfnamefont {S.}~\bibnamefont {Schiller}},\ }\bibfield  {title}
  {\bibinfo {title} {{Vibrational spectroscopy of HD+ with 2-ppb accuracy}},\
  }\bibfield  {journal} {\bibinfo  {journal} {Physical Review Letters}\
  }\textbf {\bibinfo {volume} {98}},\ \href
  {https://doi.org/10.1103/PhysRevLett.98.173002}
  {10.1103/PhysRevLett.98.173002} (\bibinfo {year} {2007})\BibitemShut
  {NoStop}%
\bibitem [{\citenamefont {Bressel}\ \emph {et~al.}(2012)\citenamefont
  {Bressel}, \citenamefont {Borodin}, \citenamefont {Shen}, \citenamefont
  {Hansen}, \citenamefont {Ernsting},\ and\ \citenamefont
  {Schiller}}]{Bressel2012}%
  \BibitemOpen
  \bibfield  {author} {\bibinfo {author} {\bibfnamefont {U.}~\bibnamefont
  {Bressel}}, \bibinfo {author} {\bibfnamefont {A.}~\bibnamefont {Borodin}},
  \bibinfo {author} {\bibfnamefont {J.}~\bibnamefont {Shen}}, \bibinfo {author}
  {\bibfnamefont {M.}~\bibnamefont {Hansen}}, \bibinfo {author} {\bibfnamefont
  {I.}~\bibnamefont {Ernsting}},\ and\ \bibinfo {author} {\bibfnamefont
  {S.}~\bibnamefont {Schiller}},\ }\bibfield  {title} {\bibinfo {title}
  {{Manipulation of Individual Hyperfine States in Cold Trapped Molecular Ions
  and Application to
  {\$}{\{}{\textbackslash}mathrm{\{}HD{\}}{\}}{\^{}}{\{}+{\}}{\$} Frequency
  Metrology}},\ }\href {https://doi.org/10.1103/PhysRevLett.108.183003}
  {\bibfield  {journal} {\bibinfo  {journal} {Physical Review Letters}\
  }\textbf {\bibinfo {volume} {108}},\ \bibinfo {pages} {183003} (\bibinfo
  {year} {2012})}\BibitemShut {NoStop}%
\bibitem [{\citenamefont {Dickenson}\ \emph {et~al.}(2013)\citenamefont
  {Dickenson}, \citenamefont {Niu}, \citenamefont {Salumbides}, \citenamefont
  {Komasa}, \citenamefont {Eikema}, \citenamefont {Pachucki},\ and\
  \citenamefont {Ubachs}}]{Dickenson2013}%
  \BibitemOpen
  \bibfield  {author} {\bibinfo {author} {\bibfnamefont {G.~D.}\ \bibnamefont
  {Dickenson}}, \bibinfo {author} {\bibfnamefont {M.~L.}\ \bibnamefont {Niu}},
  \bibinfo {author} {\bibfnamefont {E.~J.}\ \bibnamefont {Salumbides}},
  \bibinfo {author} {\bibfnamefont {J.}~\bibnamefont {Komasa}}, \bibinfo
  {author} {\bibfnamefont {K.~S.~E.}\ \bibnamefont {Eikema}}, \bibinfo {author}
  {\bibfnamefont {K.}~\bibnamefont {Pachucki}},\ and\ \bibinfo {author}
  {\bibfnamefont {W.}~\bibnamefont {Ubachs}},\ }\bibfield  {title} {\bibinfo
  {title} {{Fundamental Vibration of Molecular Hydrogen}},\ }\href
  {https://doi.org/10.1103/PhysRevLett.110.193601} {\bibfield  {journal}
  {\bibinfo  {journal} {Physical Review Letters}\ }\textbf {\bibinfo {volume}
  {110}},\ \bibinfo {pages} {193601} (\bibinfo {year} {2013})}\BibitemShut
  {NoStop}%
\bibitem [{\citenamefont {Biesheuvel}\ \emph
  {et~al.}(2016{\natexlab{a}})\citenamefont {Biesheuvel}, \citenamefont {Karr},
  \citenamefont {Hilico}, \citenamefont {Eikema}, \citenamefont {Ubachs},\ and\
  \citenamefont {Koelemeij}}]{Biesheuvel2016a}%
  \BibitemOpen
  \bibfield  {author} {\bibinfo {author} {\bibfnamefont {J.}~\bibnamefont
  {Biesheuvel}}, \bibinfo {author} {\bibfnamefont {J.-P.}\ \bibnamefont
  {Karr}}, \bibinfo {author} {\bibfnamefont {L.}~\bibnamefont {Hilico}},
  \bibinfo {author} {\bibfnamefont {K.~S.~E.}\ \bibnamefont {Eikema}}, \bibinfo
  {author} {\bibfnamefont {W.}~\bibnamefont {Ubachs}},\ and\ \bibinfo {author}
  {\bibfnamefont {J.~C.~J.}\ \bibnamefont {Koelemeij}},\ }\bibfield  {title}
  {\bibinfo {title} {{Probing QED and fundamental constants through laser
  spectroscopy of vibrational transitions in HD+}},\ }\href
  {https://doi.org/10.1038/ncomms10385} {\bibfield  {journal} {\bibinfo
  {journal} {Nature Communications}\ }\textbf {\bibinfo {volume} {7}},\
  \bibinfo {pages} {10385} (\bibinfo {year} {2016}{\natexlab{a}})}\BibitemShut
  {NoStop}%
\bibitem [{\citenamefont {Biesheuvel}\ \emph
  {et~al.}(2016{\natexlab{b}})\citenamefont {Biesheuvel}, \citenamefont {Karr},
  \citenamefont {Hilico}, \citenamefont {Eikema}, \citenamefont {Ubachs},\ and\
  \citenamefont {Koelemeij}}]{Biesheuvel2016b}%
  \BibitemOpen
  \bibfield  {author} {\bibinfo {author} {\bibfnamefont {J.}~\bibnamefont
  {Biesheuvel}}, \bibinfo {author} {\bibfnamefont {J.-P.}\ \bibnamefont
  {Karr}}, \bibinfo {author} {\bibfnamefont {L.}~\bibnamefont {Hilico}},
  \bibinfo {author} {\bibfnamefont {K.~S.~E.}\ \bibnamefont {Eikema}}, \bibinfo
  {author} {\bibfnamefont {W.}~\bibnamefont {Ubachs}},\ and\ \bibinfo {author}
  {\bibfnamefont {J.~C.~J.}\ \bibnamefont {Koelemeij}},\ }\bibfield  {title}
  {\bibinfo {title} {{High-precision spectroscopy of the HD+ molecule at the
  1-p.p.b. level}},\ }\href {https://doi.org/10.1007/s00340-016-6576-8}
  {\bibfield  {journal} {\bibinfo  {journal} {Applied Physics B}\ }\textbf
  {\bibinfo {volume} {123}},\ \bibinfo {pages} {23} (\bibinfo {year}
  {2016}{\natexlab{b}})}\BibitemShut {NoStop}%
\bibitem [{\citenamefont {Korobov}\ \emph {et~al.}(2017)\citenamefont
  {Korobov}, \citenamefont {Hilico},\ and\ \citenamefont {Karr}}]{Korobov2017}%
  \BibitemOpen
  \bibfield  {author} {\bibinfo {author} {\bibfnamefont {V.~I.}\ \bibnamefont
  {Korobov}}, \bibinfo {author} {\bibfnamefont {L.}~\bibnamefont {Hilico}},\
  and\ \bibinfo {author} {\bibfnamefont {J.-P.}\ \bibnamefont {Karr}},\
  }\bibfield  {title} {\bibinfo {title} {{Fundamental Transitions and
  Ionization Energies of the Hydrogen Molecular Ions with Few ppt
  Uncertainty}},\ }\href {https://doi.org/10.1103/PhysRevLett.118.233001}
  {\bibfield  {journal} {\bibinfo  {journal} {Physical Review Letters}\
  }\textbf {\bibinfo {volume} {118}},\ \bibinfo {pages} {233001} (\bibinfo
  {year} {2017})}\BibitemShut {NoStop}%
\bibitem [{\citenamefont {Aznabayev}\ \emph {et~al.}(2019)\citenamefont
  {Aznabayev}, \citenamefont {Bekbaev},\ and\ \citenamefont
  {Korobov}}]{Aznabayev2019}%
  \BibitemOpen
  \bibfield  {author} {\bibinfo {author} {\bibfnamefont {D.~T.}\ \bibnamefont
  {Aznabayev}}, \bibinfo {author} {\bibfnamefont {A.~K.}\ \bibnamefont
  {Bekbaev}},\ and\ \bibinfo {author} {\bibfnamefont {V.~I.}\ \bibnamefont
  {Korobov}},\ }\bibfield  {title} {\bibinfo {title} {{Leading-order
  relativistic corrections to the rovibrational spectrum of
  {\$}{\{}{\textbackslash}text{\{}H{\}}{\}}{\_}{\{}2{\}}{\{}{\textbackslash}phantom{\{}{\textbackslash}rule{\{}0.16em{\}}{\{}0ex{\}}{\}}{\}}{\^{}}{\{}+{\}}{\$}
  and {\$}{\{}{\textbackslash}mathrm{\{}HD{\}}{\}}{\^{}}{\{}+{\}}{\$} molecular
  ions}},\ }\href {https://doi.org/10.1103/PhysRevA.99.012501} {\bibfield
  {journal} {\bibinfo  {journal} {Physical Review A}\ }\textbf {\bibinfo
  {volume} {99}},\ \bibinfo {pages} {12501} (\bibinfo {year}
  {2019})}\BibitemShut {NoStop}%
\bibitem [{\citenamefont {Fink}\ and\ \citenamefont {Myers}(2020)}]{Fink2020}%
  \BibitemOpen
  \bibfield  {author} {\bibinfo {author} {\bibfnamefont {D.~J.}\ \bibnamefont
  {Fink}}\ and\ \bibinfo {author} {\bibfnamefont {E.~G.}\ \bibnamefont
  {Myers}},\ }\bibfield  {title} {\bibinfo {title} {{Deuteron-to-Proton Mass
  Ratio from the Cyclotron Frequency Ratio of
  {\$}{\{}{\textbackslash}mathrm{\{}H{\}}{\}}{\_}{\{}2{\}}{\^{}}{\{}+{\}}{\$}
  to {\$}{\{}{\textbackslash}mathrm{\{}D{\}}{\}}{\^{}}{\{}+{\}}{\$} with
  {\$}{\{}{\textbackslash}mathrm{\{}H{\}}{\}}{\_}{\{}2{\}}{\^{}}{\{}+{\}}{\$}
  in a Resolved Vibrational State}},\ }\href
  {https://doi.org/10.1103/PhysRevLett.124.013001} {\bibfield  {journal}
  {\bibinfo  {journal} {Physical Review Letters}\ }\textbf {\bibinfo {volume}
  {124}},\ \bibinfo {pages} {13001} (\bibinfo {year} {2020})}\BibitemShut
  {NoStop}%
\bibitem [{\citenamefont {Rau}\ \emph {et~al.}(2020)\citenamefont {Rau},
  \citenamefont {Hei{\ss}e}, \citenamefont {K{\"{o}}hler-Langes}, \citenamefont
  {Sasidharan}, \citenamefont {Haas}, \citenamefont {Renisch}, \citenamefont
  {D{\"{u}}llmann}, \citenamefont {Quint}, \citenamefont {Sturm},\ and\
  \citenamefont {Blaum}}]{Rau2020}%
  \BibitemOpen
  \bibfield  {author} {\bibinfo {author} {\bibfnamefont {S.}~\bibnamefont
  {Rau}}, \bibinfo {author} {\bibfnamefont {F.}~\bibnamefont {Hei{\ss}e}},
  \bibinfo {author} {\bibfnamefont {F.}~\bibnamefont {K{\"{o}}hler-Langes}},
  \bibinfo {author} {\bibfnamefont {S.}~\bibnamefont {Sasidharan}}, \bibinfo
  {author} {\bibfnamefont {R.}~\bibnamefont {Haas}}, \bibinfo {author}
  {\bibfnamefont {D.}~\bibnamefont {Renisch}}, \bibinfo {author} {\bibfnamefont
  {C.~E.}\ \bibnamefont {D{\"{u}}llmann}}, \bibinfo {author} {\bibfnamefont
  {W.}~\bibnamefont {Quint}}, \bibinfo {author} {\bibfnamefont
  {S.}~\bibnamefont {Sturm}},\ and\ \bibinfo {author} {\bibfnamefont
  {K.}~\bibnamefont {Blaum}},\ }\bibfield  {title} {\bibinfo {title} {{Penning
  trap mass measurements of the deuteron and the HD+ molecular ion}},\ }\href
  {https://doi.org/10.1038/s41586-020-2628-7} {\bibfield  {journal} {\bibinfo
  {journal} {Nature}\ }\textbf {\bibinfo {volume} {585}},\ \bibinfo {pages}
  {43} (\bibinfo {year} {2020})}\BibitemShut {NoStop}%
\bibitem [{\citenamefont {Germann}\ \emph {et~al.}(2021)\citenamefont
  {Germann}, \citenamefont {Patra}, \citenamefont {Karr}, \citenamefont
  {Hilico}, \citenamefont {Korobov}, \citenamefont {Salumbides}, \citenamefont
  {Eikema}, \citenamefont {Ubachs},\ and\ \citenamefont
  {Koelemeij}}]{Germann2021}%
  \BibitemOpen
  \bibfield  {author} {\bibinfo {author} {\bibfnamefont {M.}~\bibnamefont
  {Germann}}, \bibinfo {author} {\bibfnamefont {S.}~\bibnamefont {Patra}},
  \bibinfo {author} {\bibfnamefont {J.~P.}\ \bibnamefont {Karr}}, \bibinfo
  {author} {\bibfnamefont {L.}~\bibnamefont {Hilico}}, \bibinfo {author}
  {\bibfnamefont {V.~I.}\ \bibnamefont {Korobov}}, \bibinfo {author}
  {\bibfnamefont {E.~J.}\ \bibnamefont {Salumbides}}, \bibinfo {author}
  {\bibfnamefont {K.~S.}\ \bibnamefont {Eikema}}, \bibinfo {author}
  {\bibfnamefont {W.}~\bibnamefont {Ubachs}},\ and\ \bibinfo {author}
  {\bibfnamefont {J.~C.}\ \bibnamefont {Koelemeij}},\ }\bibfield  {title}
  {\bibinfo {title} {{Three-body QED test and fifth-force constraint from
  vibrations and rotations of HD+}},\ }\bibfield  {journal} {\bibinfo
  {journal} {Physical Review Research}\ }\textbf {\bibinfo {volume} {3}},\
  \href {https://doi.org/10.1103/PhysRevResearch.3.L022028}
  {10.1103/PhysRevResearch.3.L022028} (\bibinfo {year} {2021})\BibitemShut
  {NoStop}%
\bibitem [{\citenamefont {Kortunov}\ \emph {et~al.}(2021)\citenamefont
  {Kortunov}, \citenamefont {Alighanbari}, \citenamefont {Hansen},
  \citenamefont {Giri}, \citenamefont {Korobov},\ and\ \citenamefont
  {Schiller}}]{Kortunov2021}%
  \BibitemOpen
  \bibfield  {author} {\bibinfo {author} {\bibfnamefont {I.~V.}\ \bibnamefont
  {Kortunov}}, \bibinfo {author} {\bibfnamefont {S.}~\bibnamefont
  {Alighanbari}}, \bibinfo {author} {\bibfnamefont {M.~G.}\ \bibnamefont
  {Hansen}}, \bibinfo {author} {\bibfnamefont {G.~S.}\ \bibnamefont {Giri}},
  \bibinfo {author} {\bibfnamefont {V.~I.}\ \bibnamefont {Korobov}},\ and\
  \bibinfo {author} {\bibfnamefont {S.}~\bibnamefont {Schiller}},\ }\bibfield
  {title} {\bibinfo {title} {{Proton–electron mass ratio by high-resolution
  optical spectroscopy of ion ensembles in the resolved-carrier regime}},\
  }\href {https://doi.org/10.1038/s41567-020-01150-7} {\bibfield  {journal}
  {\bibinfo  {journal} {Nature Physics}\ }\textbf {\bibinfo {volume} {17}},\
  \bibinfo {pages} {569} (\bibinfo {year} {2021})}\BibitemShut {NoStop}%
\bibitem [{\citenamefont {Alighanbari}\ \emph {et~al.}(2023)\citenamefont
  {Alighanbari}, \citenamefont {Kortunov}, \citenamefont {Giri},\ and\
  \citenamefont {Schiller}}]{Alighanbari2023}%
  \BibitemOpen
  \bibfield  {author} {\bibinfo {author} {\bibfnamefont {S.}~\bibnamefont
  {Alighanbari}}, \bibinfo {author} {\bibfnamefont {I.~V.}\ \bibnamefont
  {Kortunov}}, \bibinfo {author} {\bibfnamefont {G.~S.}\ \bibnamefont {Giri}},\
  and\ \bibinfo {author} {\bibfnamefont {S.}~\bibnamefont {Schiller}},\
  }\bibfield  {title} {\bibinfo {title} {{Test of charged baryon interaction
  with high-resolution vibrational spectroscopy of molecular hydrogen ions}},\
  }\href {https://doi.org/10.1038/s41567-023-02088-2} {\bibfield  {journal}
  {\bibinfo  {journal} {Nature Physics}\ }\textbf {\bibinfo {volume} {19}},\
  \bibinfo {pages} {1263} (\bibinfo {year} {2023})}\BibitemShut {NoStop}%
\bibitem [{\citenamefont {Komasa}\ \emph {et~al.}(2011)\citenamefont {Komasa},
  \citenamefont {Piszczatowski}, \citenamefont {{\L}ach}, \citenamefont
  {Przybytek}, \citenamefont {Jeziorski},\ and\ \citenamefont
  {Pachucki}}]{Komasa2011}%
  \BibitemOpen
  \bibfield  {author} {\bibinfo {author} {\bibfnamefont {J.}~\bibnamefont
  {Komasa}}, \bibinfo {author} {\bibfnamefont {K.}~\bibnamefont
  {Piszczatowski}}, \bibinfo {author} {\bibfnamefont {G.}~\bibnamefont
  {{\L}ach}}, \bibinfo {author} {\bibfnamefont {M.}~\bibnamefont {Przybytek}},
  \bibinfo {author} {\bibfnamefont {B.}~\bibnamefont {Jeziorski}},\ and\
  \bibinfo {author} {\bibfnamefont {K.}~\bibnamefont {Pachucki}},\ }\bibfield
  {title} {\bibinfo {title} {{Quantum Electrodynamics Effects in Rovibrational
  Spectra of Molecular Hydrogen}},\ }\href {https://doi.org/10.1021/ct200438t}
  {\bibfield  {journal} {\bibinfo  {journal} {Journal of Chemical Theory and
  Computation}\ }\textbf {\bibinfo {volume} {7}},\ \bibinfo {pages} {3105}
  (\bibinfo {year} {2011})}\BibitemShut {NoStop}%
\bibitem [{\citenamefont {Puchalski}\ \emph {et~al.}(2019)\citenamefont
  {Puchalski}, \citenamefont {Komasa}, \citenamefont {Czachorowski},\ and\
  \citenamefont {Pachucki}}]{Puchalski2019}%
  \BibitemOpen
  \bibfield  {author} {\bibinfo {author} {\bibfnamefont {M.}~\bibnamefont
  {Puchalski}}, \bibinfo {author} {\bibfnamefont {J.}~\bibnamefont {Komasa}},
  \bibinfo {author} {\bibfnamefont {P.}~\bibnamefont {Czachorowski}},\ and\
  \bibinfo {author} {\bibfnamefont {K.}~\bibnamefont {Pachucki}},\ }\bibfield
  {title} {\bibinfo {title} {{Nonadiabatic QED Correction to the Dissociation
  Energy of the Hydrogen Molecule}},\ }\href
  {https://doi.org/10.1103/PhysRevLett.122.103003} {\bibfield  {journal}
  {\bibinfo  {journal} {Physical Review Letters}\ }\textbf {\bibinfo {volume}
  {122}},\ \bibinfo {pages} {103003} (\bibinfo {year} {2019})}\BibitemShut
  {NoStop}%
\bibitem [{\citenamefont {H{\"{o}}lsch}\ \emph {et~al.}(2019)\citenamefont
  {H{\"{o}}lsch}, \citenamefont {Beyer}, \citenamefont {Salumbides},
  \citenamefont {Eikema}, \citenamefont {Ubachs}, \citenamefont {Jungen},\ and\
  \citenamefont {Merkt}}]{Hoelsch2019}%
  \BibitemOpen
  \bibfield  {author} {\bibinfo {author} {\bibfnamefont {N.}~\bibnamefont
  {H{\"{o}}lsch}}, \bibinfo {author} {\bibfnamefont {M.}~\bibnamefont {Beyer}},
  \bibinfo {author} {\bibfnamefont {E.~J.}\ \bibnamefont {Salumbides}},
  \bibinfo {author} {\bibfnamefont {K.~S.}\ \bibnamefont {Eikema}}, \bibinfo
  {author} {\bibfnamefont {W.}~\bibnamefont {Ubachs}}, \bibinfo {author}
  {\bibfnamefont {C.}~\bibnamefont {Jungen}},\ and\ \bibinfo {author}
  {\bibfnamefont {F.}~\bibnamefont {Merkt}},\ }\bibfield  {title} {\bibinfo
  {title} {{Benchmarking Theory with an Improved Measurement of the Ionization
  and Dissociation Energies of
  {\$}{\{}{\textbackslash}mathrm{\{}H{\}}{\}}{\_}{\{}2{\}}{\$}}},\ }\href
  {https://doi.org/10.1103/PhysRevLett.122.103002} {\bibfield  {journal}
  {\bibinfo  {journal} {Physical Review Letters}\ }\textbf {\bibinfo {volume}
  {122}},\ \bibinfo {pages} {103002} (\bibinfo {year} {2019})}\BibitemShut
  {NoStop}%
\bibitem [{\citenamefont {Dulick}\ \emph {et~al.}(1998)\citenamefont {Dulick},
  \citenamefont {Zhang}, \citenamefont {Guo},\ and\ \citenamefont
  {Bernath}}]{Dulick1998LiH}%
  \BibitemOpen
  \bibfield  {author} {\bibinfo {author} {\bibfnamefont {M.}~\bibnamefont
  {Dulick}}, \bibinfo {author} {\bibfnamefont {K.-Q.}\ \bibnamefont {Zhang}},
  \bibinfo {author} {\bibfnamefont {B.}~\bibnamefont {Guo}},\ and\ \bibinfo
  {author} {\bibfnamefont {P.~F.}\ \bibnamefont {Bernath}},\ }\bibfield
  {title} {\bibinfo {title} {{Far- and Mid-Infrared Emission Spectroscopy of
  LiH and LiD}},\ }\href
  {https://doi.org/https://doi.org/10.1006/jmsp.1997.7430} {\bibfield
  {journal} {\bibinfo  {journal} {Journal of Molecular Spectroscopy}\ }\textbf
  {\bibinfo {volume} {188}},\ \bibinfo {pages} {14} (\bibinfo {year}
  {1998})}\BibitemShut {NoStop}%
\bibitem [{\citenamefont {Markus}\ \emph {et~al.}(2016)\citenamefont {Markus},
  \citenamefont {Hodges}, \citenamefont {Perry}, \citenamefont {Kocheril},
  \citenamefont {M{\"{u}}ller},\ and\ \citenamefont {McCall}}]{Markus2016OH+}%
  \BibitemOpen
  \bibfield  {author} {\bibinfo {author} {\bibfnamefont {C.~R.}\ \bibnamefont
  {Markus}}, \bibinfo {author} {\bibfnamefont {J.~N.}\ \bibnamefont {Hodges}},
  \bibinfo {author} {\bibfnamefont {A.~J.}\ \bibnamefont {Perry}}, \bibinfo
  {author} {\bibfnamefont {G.~S.}\ \bibnamefont {Kocheril}}, \bibinfo {author}
  {\bibfnamefont {H.~S.~P.}\ \bibnamefont {M{\"{u}}ller}},\ and\ \bibinfo
  {author} {\bibfnamefont {B.~J.}\ \bibnamefont {McCall}},\ }\bibfield  {title}
  {\bibinfo {title} {{HIGH PRECISION ROVIBRATIONAL SPECTROSCOPY OF OH+}},\
  }\href {https://doi.org/10.3847/0004-637X/817/2/138} {\bibfield  {journal}
  {\bibinfo  {journal} {The Astrophysical Journal}\ }\textbf {\bibinfo {volume}
  {817}},\ \bibinfo {pages} {138} (\bibinfo {year} {2016})}\BibitemShut
  {NoStop}%
\bibitem [{\citenamefont {McKellar}\ \emph {et~al.}(1976)\citenamefont
  {McKellar}, \citenamefont {Goetz},\ and\ \citenamefont
  {Ramsay}}]{McKellar1976}%
  \BibitemOpen
  \bibfield  {author} {\bibinfo {author} {\bibfnamefont {A.~R.~W.}\
  \bibnamefont {McKellar}}, \bibinfo {author} {\bibfnamefont {W.}~\bibnamefont
  {Goetz}},\ and\ \bibinfo {author} {\bibfnamefont {D.~A.}\ \bibnamefont
  {Ramsay}},\ }\bibfield  {title} {\bibinfo {title} {{The rotation-vibration
  spectrum of HD-Wavelength and intensity measurements of the 3-0, 4-0, 5-0,
  and 6-0 electric dipole bands}},\ }\href@noop {} {\bibfield  {journal}
  {\bibinfo  {journal} {Astrophysical Journal, vol. 207, July 15, 1976, pt. 1,
  p. 663-670.}\ }\textbf {\bibinfo {volume} {207}},\ \bibinfo {pages} {663}
  (\bibinfo {year} {1976})}\BibitemShut {NoStop}%
\bibitem [{\citenamefont {WONG}(1994)}]{Wong1994DeuteronMatterRadius}%
  \BibitemOpen
  \bibfield  {author} {\bibinfo {author} {\bibfnamefont {C.~W.}\ \bibnamefont
  {WONG}},\ }\bibfield  {title} {\bibinfo {title} {{DEUTERON RADIUS AND NUCLEAR
  FORCES IN FREE SPACE}},\ }\href {https://doi.org/10.1142/S0218301394000255}
  {\bibfield  {journal} {\bibinfo  {journal} {International Journal of Modern
  Physics E}\ }\textbf {\bibinfo {volume} {03}},\ \bibinfo {pages} {821}
  (\bibinfo {year} {1994})}\BibitemShut {NoStop}%
\bibitem [{\citenamefont {Casten}(2000)}]{CastenBook}%
  \BibitemOpen
  \bibfield  {author} {\bibinfo {author} {\bibfnamefont {R.~F.}\ \bibnamefont
  {Casten}},\ }\href@noop {} {\emph {\bibinfo {title} {{Nuclear structure from
  a simple perspective}}}},\ Vol.~\bibinfo {volume} {23}\ (\bibinfo
  {publisher} {Oxford University Press on Demand},\ \bibinfo {year}
  {2000})\BibitemShut {NoStop}%
\bibitem [{\citenamefont {Igo}\ \emph {et~al.}(1979)\citenamefont {Igo},
  \citenamefont {Adams}, \citenamefont {Bauer}, \citenamefont {Pauletta},
  \citenamefont {Whitten}, \citenamefont {Wreikat}, \citenamefont {Hoffmann},
  \citenamefont {Blanpied}, \citenamefont {Coker}, \citenamefont {Harvey},
  \citenamefont {Liljestrand}, \citenamefont {Ray}, \citenamefont {Spencer},
  \citenamefont {Thiessen}, \citenamefont {Glashausser}, \citenamefont {Hintz},
  \citenamefont {Oothoudt}, \citenamefont {Nann}, \citenamefont {Seth},
  \citenamefont {Wood}, \citenamefont {McDaniels},\ and\ \citenamefont
  {Gazzaly}}]{Igo1979}%
  \BibitemOpen
  \bibfield  {author} {\bibinfo {author} {\bibfnamefont {G.}~\bibnamefont
  {Igo}}, \bibinfo {author} {\bibfnamefont {G.~S.}\ \bibnamefont {Adams}},
  \bibinfo {author} {\bibfnamefont {T.~S.}\ \bibnamefont {Bauer}}, \bibinfo
  {author} {\bibfnamefont {G.}~\bibnamefont {Pauletta}}, \bibinfo {author}
  {\bibfnamefont {C.~A.}\ \bibnamefont {Whitten}}, \bibinfo {author}
  {\bibfnamefont {A.}~\bibnamefont {Wreikat}}, \bibinfo {author} {\bibfnamefont
  {G.~W.}\ \bibnamefont {Hoffmann}}, \bibinfo {author} {\bibfnamefont {G.~S.}\
  \bibnamefont {Blanpied}}, \bibinfo {author} {\bibfnamefont {W.~R.}\
  \bibnamefont {Coker}}, \bibinfo {author} {\bibfnamefont {C.}~\bibnamefont
  {Harvey}}, \bibinfo {author} {\bibfnamefont {R.~P.}\ \bibnamefont
  {Liljestrand}}, \bibinfo {author} {\bibfnamefont {L.}~\bibnamefont {Ray}},
  \bibinfo {author} {\bibfnamefont {J.~E.}\ \bibnamefont {Spencer}}, \bibinfo
  {author} {\bibfnamefont {H.~A.}\ \bibnamefont {Thiessen}}, \bibinfo {author}
  {\bibfnamefont {C.}~\bibnamefont {Glashausser}}, \bibinfo {author}
  {\bibfnamefont {N.~M.}\ \bibnamefont {Hintz}}, \bibinfo {author}
  {\bibfnamefont {M.~A.}\ \bibnamefont {Oothoudt}}, \bibinfo {author}
  {\bibfnamefont {H.}~\bibnamefont {Nann}}, \bibinfo {author} {\bibfnamefont
  {K.~K.}\ \bibnamefont {Seth}}, \bibinfo {author} {\bibfnamefont {B.~E.}\
  \bibnamefont {Wood}}, \bibinfo {author} {\bibfnamefont {D.~K.}\ \bibnamefont
  {McDaniels}},\ and\ \bibinfo {author} {\bibfnamefont {M.}~\bibnamefont
  {Gazzaly}},\ }\bibfield  {title} {\bibinfo {title} {{Elastic differential
  cross sections and analyzing powers for p+40,42,44,48Ca at 0.8 GeV}},\ }\href
  {https://doi.org/https://doi.org/10.1016/0370-2693(79)90510-0} {\bibfield
  {journal} {\bibinfo  {journal} {Physics Letters B}\ }\textbf {\bibinfo
  {volume} {81}},\ \bibinfo {pages} {151} (\bibinfo {year} {1979})}\BibitemShut
  {NoStop}%
\bibitem [{\citenamefont {McCamis}\ \emph {et~al.}(1986)\citenamefont
  {McCamis}, \citenamefont {Nasr}, \citenamefont {Birchall}, \citenamefont
  {Davison}, \citenamefont {van Oers}, \citenamefont {Verheijen}, \citenamefont
  {Carlson}, \citenamefont {Cox}, \citenamefont {Clark}, \citenamefont
  {Cooper}, \citenamefont {Hama},\ and\ \citenamefont {Mercer}}]{McCamis1986}%
  \BibitemOpen
  \bibfield  {author} {\bibinfo {author} {\bibfnamefont {R.~H.}\ \bibnamefont
  {McCamis}}, \bibinfo {author} {\bibfnamefont {T.~N.}\ \bibnamefont {Nasr}},
  \bibinfo {author} {\bibfnamefont {J.}~\bibnamefont {Birchall}}, \bibinfo
  {author} {\bibfnamefont {N.~E.}\ \bibnamefont {Davison}}, \bibinfo {author}
  {\bibfnamefont {W.~T.~H.}\ \bibnamefont {van Oers}}, \bibinfo {author}
  {\bibfnamefont {P.~J.~T.}\ \bibnamefont {Verheijen}}, \bibinfo {author}
  {\bibfnamefont {R.~F.}\ \bibnamefont {Carlson}}, \bibinfo {author}
  {\bibfnamefont {A.~J.}\ \bibnamefont {Cox}}, \bibinfo {author} {\bibfnamefont
  {B.~C.}\ \bibnamefont {Clark}}, \bibinfo {author} {\bibfnamefont {E.~D.}\
  \bibnamefont {Cooper}}, \bibinfo {author} {\bibfnamefont {S.}~\bibnamefont
  {Hama}},\ and\ \bibinfo {author} {\bibfnamefont {R.~L.}\ \bibnamefont
  {Mercer}},\ }\bibfield  {title} {\bibinfo {title} {{Elastic scattering of
  protons from 40,42,44,48Ca from 20 to 50 MeV and nuclear matter radii}},\
  }\href {https://doi.org/10.1103/PhysRevC.33.1624} {\bibfield  {journal}
  {\bibinfo  {journal} {Physical Review C}\ }\textbf {\bibinfo {volume} {33}},\
  \bibinfo {pages} {1624} (\bibinfo {year} {1986})}\BibitemShut {NoStop}%
\bibitem [{\citenamefont {Zenihiro}\ \emph {et~al.}(2018)\citenamefont
  {Zenihiro}, \citenamefont {Sakaguchi}, \citenamefont {Terashima},
  \citenamefont {Uesaka}, \citenamefont {Hagen}, \citenamefont {Itoh},
  \citenamefont {Murakami}, \citenamefont {Nakatsugawa}, \citenamefont
  {Ohnishi}, \citenamefont {Sagawa}, \citenamefont {Takeda}, \citenamefont
  {Uchida}, \citenamefont {Yoshida}, \citenamefont {Yoshida},\ and\
  \citenamefont {Yosoi}}]{Zenihiro2018}%
  \BibitemOpen
  \bibfield  {author} {\bibinfo {author} {\bibfnamefont {J.}~\bibnamefont
  {Zenihiro}}, \bibinfo {author} {\bibfnamefont {H.}~\bibnamefont {Sakaguchi}},
  \bibinfo {author} {\bibfnamefont {S.}~\bibnamefont {Terashima}}, \bibinfo
  {author} {\bibfnamefont {T.}~\bibnamefont {Uesaka}}, \bibinfo {author}
  {\bibfnamefont {G.}~\bibnamefont {Hagen}}, \bibinfo {author} {\bibfnamefont
  {M.}~\bibnamefont {Itoh}}, \bibinfo {author} {\bibfnamefont {T.}~\bibnamefont
  {Murakami}}, \bibinfo {author} {\bibfnamefont {Y.}~\bibnamefont
  {Nakatsugawa}}, \bibinfo {author} {\bibfnamefont {T.}~\bibnamefont
  {Ohnishi}}, \bibinfo {author} {\bibfnamefont {H.}~\bibnamefont {Sagawa}},
  \bibinfo {author} {\bibfnamefont {H.}~\bibnamefont {Takeda}}, \bibinfo
  {author} {\bibfnamefont {M.}~\bibnamefont {Uchida}}, \bibinfo {author}
  {\bibfnamefont {H.~P.}\ \bibnamefont {Yoshida}}, \bibinfo {author}
  {\bibfnamefont {S.}~\bibnamefont {Yoshida}},\ and\ \bibinfo {author}
  {\bibfnamefont {M.}~\bibnamefont {Yosoi}},\ }\bibfield  {title} {\bibinfo
  {title} {{Direct determination of the neutron skin thicknesses in 40,48Ca
  from proton elastic scattering at Ep = 295 MeV}},\ }\href@noop {} {\bibfield
  {journal} {\bibinfo  {journal} {arXiv:1810.11796}\ } (\bibinfo {year}
  {2018})}\BibitemShut {NoStop}%
\bibitem [{\citenamefont {Tanaka}\ \emph {et~al.}(2020)\citenamefont {Tanaka},
  \citenamefont {Takechi}, \citenamefont {Homma}, \citenamefont {Fukuda},
  \citenamefont {Nishimura}, \citenamefont {Suzuki}, \citenamefont {Tanaka},
  \citenamefont {Moriguchi}, \citenamefont {Ahn}, \citenamefont {Aimaganbetov},
  \citenamefont {Amano}, \citenamefont {Arakawa}, \citenamefont {Bagchi},
  \citenamefont {Behr}, \citenamefont {Burtebayev}, \citenamefont {Chikaato},
  \citenamefont {Du}, \citenamefont {Ebata}, \citenamefont {Fujii},
  \citenamefont {Fukuda}, \citenamefont {Geissel}, \citenamefont {Hori},
  \citenamefont {Horiuchi}, \citenamefont {Hoshino}, \citenamefont {Igosawa},
  \citenamefont {Ikeda}, \citenamefont {Inabe}, \citenamefont {Inomata},
  \citenamefont {Itahashi}, \citenamefont {Izumikawa}, \citenamefont {Kamioka},
  \citenamefont {Kanda}, \citenamefont {Kato}, \citenamefont {Kenzhina},
  \citenamefont {Korkulu}, \citenamefont {Kuk}, \citenamefont {Kusaka},
  \citenamefont {Matsuta}, \citenamefont {Mihara}, \citenamefont {Miyata},
  \citenamefont {Nagae}, \citenamefont {Nakamura}, \citenamefont {Nassurlla},
  \citenamefont {Nishimuro}, \citenamefont {Nishizuka}, \citenamefont
  {Ohnishi}, \citenamefont {Ohtake}, \citenamefont {Ohtsubo}, \citenamefont
  {Omika}, \citenamefont {Ong}, \citenamefont {Ozawa}, \citenamefont
  {Prochazka}, \citenamefont {Sakurai}, \citenamefont {Scheidenberger},
  \citenamefont {Shimizu}, \citenamefont {Sugihara}, \citenamefont {Sumikama},
  \citenamefont {Suzuki}, \citenamefont {Suzuki}, \citenamefont {Takeda},
  \citenamefont {Tanaka}, \citenamefont {Tanihata}, \citenamefont {Wada},
  \citenamefont {Wakayama}, \citenamefont {Yagi}, \citenamefont {Yamaguchi},
  \citenamefont {Yanagihara}, \citenamefont {Yanagisawa}, \citenamefont
  {Yoshida},\ and\ \citenamefont {Zholdybayev}}]{Tanaka2000}%
  \BibitemOpen
  \bibfield  {author} {\bibinfo {author} {\bibfnamefont {M.}~\bibnamefont
  {Tanaka}}, \bibinfo {author} {\bibfnamefont {M.}~\bibnamefont {Takechi}},
  \bibinfo {author} {\bibfnamefont {A.}~\bibnamefont {Homma}}, \bibinfo
  {author} {\bibfnamefont {M.}~\bibnamefont {Fukuda}}, \bibinfo {author}
  {\bibfnamefont {D.}~\bibnamefont {Nishimura}}, \bibinfo {author}
  {\bibfnamefont {T.}~\bibnamefont {Suzuki}}, \bibinfo {author} {\bibfnamefont
  {Y.}~\bibnamefont {Tanaka}}, \bibinfo {author} {\bibfnamefont
  {T.}~\bibnamefont {Moriguchi}}, \bibinfo {author} {\bibfnamefont
  {D.}~\bibnamefont {Ahn}}, \bibinfo {author} {\bibfnamefont {A.}~\bibnamefont
  {Aimaganbetov}}, \bibinfo {author} {\bibfnamefont {M.}~\bibnamefont {Amano}},
  \bibinfo {author} {\bibfnamefont {H.}~\bibnamefont {Arakawa}}, \bibinfo
  {author} {\bibfnamefont {S.}~\bibnamefont {Bagchi}}, \bibinfo {author}
  {\bibfnamefont {K.-H.}\ \bibnamefont {Behr}}, \bibinfo {author}
  {\bibfnamefont {N.}~\bibnamefont {Burtebayev}}, \bibinfo {author}
  {\bibfnamefont {K.}~\bibnamefont {Chikaato}}, \bibinfo {author}
  {\bibfnamefont {H.}~\bibnamefont {Du}}, \bibinfo {author} {\bibfnamefont
  {S.}~\bibnamefont {Ebata}}, \bibinfo {author} {\bibfnamefont
  {T.}~\bibnamefont {Fujii}}, \bibinfo {author} {\bibfnamefont
  {N.}~\bibnamefont {Fukuda}}, \bibinfo {author} {\bibfnamefont
  {H.}~\bibnamefont {Geissel}}, \bibinfo {author} {\bibfnamefont
  {T.}~\bibnamefont {Hori}}, \bibinfo {author} {\bibfnamefont {W.}~\bibnamefont
  {Horiuchi}}, \bibinfo {author} {\bibfnamefont {S.}~\bibnamefont {Hoshino}},
  \bibinfo {author} {\bibfnamefont {R.}~\bibnamefont {Igosawa}}, \bibinfo
  {author} {\bibfnamefont {A.}~\bibnamefont {Ikeda}}, \bibinfo {author}
  {\bibfnamefont {N.}~\bibnamefont {Inabe}}, \bibinfo {author} {\bibfnamefont
  {K.}~\bibnamefont {Inomata}}, \bibinfo {author} {\bibfnamefont
  {K.}~\bibnamefont {Itahashi}}, \bibinfo {author} {\bibfnamefont
  {T.}~\bibnamefont {Izumikawa}}, \bibinfo {author} {\bibfnamefont
  {D.}~\bibnamefont {Kamioka}}, \bibinfo {author} {\bibfnamefont
  {N.}~\bibnamefont {Kanda}}, \bibinfo {author} {\bibfnamefont
  {I.}~\bibnamefont {Kato}}, \bibinfo {author} {\bibfnamefont {I.}~\bibnamefont
  {Kenzhina}}, \bibinfo {author} {\bibfnamefont {Z.}~\bibnamefont {Korkulu}},
  \bibinfo {author} {\bibfnamefont {Y.}~\bibnamefont {Kuk}}, \bibinfo {author}
  {\bibfnamefont {K.}~\bibnamefont {Kusaka}}, \bibinfo {author} {\bibfnamefont
  {K.}~\bibnamefont {Matsuta}}, \bibinfo {author} {\bibfnamefont
  {M.}~\bibnamefont {Mihara}}, \bibinfo {author} {\bibfnamefont
  {E.}~\bibnamefont {Miyata}}, \bibinfo {author} {\bibfnamefont
  {D.}~\bibnamefont {Nagae}}, \bibinfo {author} {\bibfnamefont
  {S.}~\bibnamefont {Nakamura}}, \bibinfo {author} {\bibfnamefont
  {M.}~\bibnamefont {Nassurlla}}, \bibinfo {author} {\bibfnamefont
  {K.}~\bibnamefont {Nishimuro}}, \bibinfo {author} {\bibfnamefont
  {K.}~\bibnamefont {Nishizuka}}, \bibinfo {author} {\bibfnamefont
  {K.}~\bibnamefont {Ohnishi}}, \bibinfo {author} {\bibfnamefont
  {M.}~\bibnamefont {Ohtake}}, \bibinfo {author} {\bibfnamefont
  {T.}~\bibnamefont {Ohtsubo}}, \bibinfo {author} {\bibfnamefont
  {S.}~\bibnamefont {Omika}}, \bibinfo {author} {\bibfnamefont
  {H.}~\bibnamefont {Ong}}, \bibinfo {author} {\bibfnamefont {A.}~\bibnamefont
  {Ozawa}}, \bibinfo {author} {\bibfnamefont {A.}~\bibnamefont {Prochazka}},
  \bibinfo {author} {\bibfnamefont {H.}~\bibnamefont {Sakurai}}, \bibinfo
  {author} {\bibfnamefont {C.}~\bibnamefont {Scheidenberger}}, \bibinfo
  {author} {\bibfnamefont {Y.}~\bibnamefont {Shimizu}}, \bibinfo {author}
  {\bibfnamefont {T.}~\bibnamefont {Sugihara}}, \bibinfo {author}
  {\bibfnamefont {T.}~\bibnamefont {Sumikama}}, \bibinfo {author}
  {\bibfnamefont {H.}~\bibnamefont {Suzuki}}, \bibinfo {author} {\bibfnamefont
  {S.}~\bibnamefont {Suzuki}}, \bibinfo {author} {\bibfnamefont
  {H.}~\bibnamefont {Takeda}}, \bibinfo {author} {\bibfnamefont
  {Y.}~\bibnamefont {Tanaka}}, \bibinfo {author} {\bibfnamefont
  {I.}~\bibnamefont {Tanihata}}, \bibinfo {author} {\bibfnamefont
  {T.}~\bibnamefont {Wada}}, \bibinfo {author} {\bibfnamefont {K.}~\bibnamefont
  {Wakayama}}, \bibinfo {author} {\bibfnamefont {S.}~\bibnamefont {Yagi}},
  \bibinfo {author} {\bibfnamefont {T.}~\bibnamefont {Yamaguchi}}, \bibinfo
  {author} {\bibfnamefont {R.}~\bibnamefont {Yanagihara}}, \bibinfo {author}
  {\bibfnamefont {Y.}~\bibnamefont {Yanagisawa}}, \bibinfo {author}
  {\bibfnamefont {K.}~\bibnamefont {Yoshida}},\ and\ \bibinfo {author}
  {\bibfnamefont {T.}~\bibnamefont {Zholdybayev}},\ }\bibfield  {title}
  {\bibinfo {title} {{Swelling of Doubly Magic 48Ca Core in Ca Isotopes beyond
  N=28}},\ }\href {https://doi.org/10.1103/PhysRevLett.124.102501} {\bibfield
  {journal} {\bibinfo  {journal} {Physical Review Letters}\ }\textbf {\bibinfo
  {volume} {124}},\ \bibinfo {pages} {102501} (\bibinfo {year}
  {2020})}\BibitemShut {NoStop}%
\bibitem [{\citenamefont {Birkhan}\ \emph {et~al.}(2017)\citenamefont
  {Birkhan}, \citenamefont {Miorelli}, \citenamefont {Bacca}, \citenamefont
  {Bassauer}, \citenamefont {Bertulani}, \citenamefont {Hagen}, \citenamefont
  {Matsubara}, \citenamefont {von Neumann-Cosel}, \citenamefont {Papenbrock},
  \citenamefont {Pietralla}, \citenamefont {Ponomarev}, \citenamefont
  {Richter}, \citenamefont {Schwenk},\ and\ \citenamefont
  {Tamii}}]{Birkhan2017}%
  \BibitemOpen
  \bibfield  {author} {\bibinfo {author} {\bibfnamefont {J.}~\bibnamefont
  {Birkhan}}, \bibinfo {author} {\bibfnamefont {M.}~\bibnamefont {Miorelli}},
  \bibinfo {author} {\bibfnamefont {S.}~\bibnamefont {Bacca}}, \bibinfo
  {author} {\bibfnamefont {S.}~\bibnamefont {Bassauer}}, \bibinfo {author}
  {\bibfnamefont {C.}~\bibnamefont {Bertulani}}, \bibinfo {author}
  {\bibfnamefont {G.}~\bibnamefont {Hagen}}, \bibinfo {author} {\bibfnamefont
  {H.}~\bibnamefont {Matsubara}}, \bibinfo {author} {\bibfnamefont
  {P.}~\bibnamefont {von Neumann-Cosel}}, \bibinfo {author} {\bibfnamefont
  {T.}~\bibnamefont {Papenbrock}}, \bibinfo {author} {\bibfnamefont
  {N.}~\bibnamefont {Pietralla}}, \bibinfo {author} {\bibfnamefont {V.~Y.}\
  \bibnamefont {Ponomarev}}, \bibinfo {author} {\bibfnamefont {A.}~\bibnamefont
  {Richter}}, \bibinfo {author} {\bibfnamefont {A.}~\bibnamefont {Schwenk}},\
  and\ \bibinfo {author} {\bibfnamefont {A.}~\bibnamefont {Tamii}},\ }\bibfield
   {title} {\bibinfo {title} {{Electric Dipole Polarizability of
  {\$}{\^{}}{\{}48{\}}{\textbackslash}mathrm{\{}Ca{\}}{\$} and Implications for
  the Neutron Skin}},\ }\href {https://doi.org/10.1103/PhysRevLett.118.252501}
  {\bibfield  {journal} {\bibinfo  {journal} {Physical Review Letters}\
  }\textbf {\bibinfo {volume} {118}},\ \bibinfo {pages} {252501} (\bibinfo
  {year} {2017})}\BibitemShut {NoStop}%
\end{thebibliography}%

\end{document}